\documentclass[12pt,preprint]{aastex}
\usepackage{bm}
\begin{document}
\title{Luminous Red Galaxy Halo Density Field Reconstruction and Application to Large Scale Structure Measurements}
%short title: LRG Halo Density Field Reconstruction
\author{Beth A. Reid}
\affil{Department of Physics}
\affil{Department of Astrophysical Sciences}
\affil{Princeton University, Princeton, NJ 08544}
\email{breid@princeton.edu}
\author{David N. Spergel}
\affil{Department of Astrophysical Sciences}
\affil{Princeton Center for Theoretical Sciences}
\affil{Princeton University, Princeton, NJ 08544}
\email{dns@astro.princeton.edu}
\and
\author{Paul Bode}
\affil{Department of Astrophysical Sciences}
\affil{Princeton University, Princeton, NJ 08544}
\email{bode@astro.princeton.edu}
\begin{abstract}
The complex relationship between the galaxy density field and the underlying matter field limits our ability to extract cosmological constraints from galaxy redshift surveys.   Our approach is to use halos rather than galaxies to trace the underlying mass distribution.  We identify Fingers-of-God (FOGs) and replace multiple galaxies in each FOG with a single halo object.  This removes the nonlinear contributions of satellite galaxies, the one-halo term.  We test our method on a large set of high-fidelity mock SDSS Luminous Red Galaxy (LRG) catalogs.  We find that the aggressive FOG compression algorithm adopted in the LRG $P(k)$ analysis of \citet{tegmark/etal:2006} leads to a $\sim 10\%$ correction to the underlying matter power spectrum at $k = 0.1 \; h \; {\rm Mpc}^{-1}$ and $\sim 40\%$ correction at $k=0.2 \; h \; {\rm Mpc}^{-1}$, thereby compromising the cosmological constraints.  In contrast, the power spectrum of our reconstructed halo density field deviates from the underlying matter power spectrum at the $\leq 1\%$ level for $k \leq 0.1 \; h \; {\rm Mpc}^{-1}$ and $\leq 4\%$ at $k=0.2 \; h \; {\rm Mpc}^{-1}$.  The reconstructed halo density field also removes the bias in the measurement of the redshift space distortion parameter $\beta$ induced by the FOG smearing of the linear redshift space distortions.\\
\end{abstract}
\section{Introduction}
\begin{figure}
\includegraphics*[scale=0.75,angle=0]{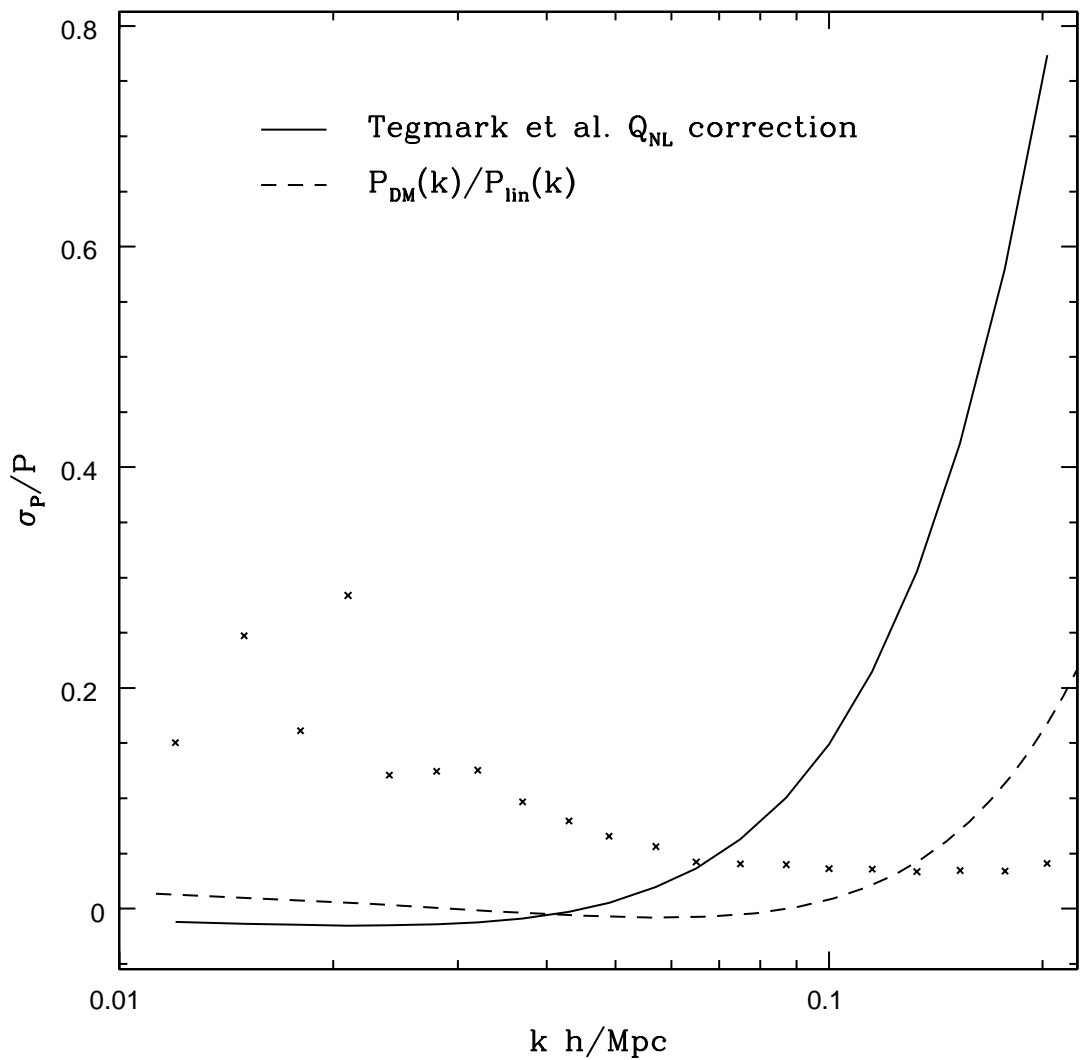}
\caption{\label{fig:qnlplotnew} Points show the error bars divided by the bandpower for the LRG power spectrum published in \citet{tegmark/etal:2006}.  The solid curve shows $(1 + Q_{NL}k^2)/(1+1.4k) - 1$, the fractional amplitude of the nonlinear $Q_{NL}$ correction to the linear power spectrum for $Q_{NL} = 31$, the best fit value in \citet{tegmark/etal:2006}.  For comparison, the dashed curve shows the smooth nonlinear correction fit to the dark matter power spectrum of Fig.~\ref{fig:DMnonlinear3}.  This is the size of the correction for our reconstructed halo density field.}
\end{figure}

Galaxy power spectra are an essential element in testing our cosmological models and studying the initial conditions of our universe.  As surveys continue to grow in size, the statistical error bars on the galaxy power spectra continue to shrink.  However, our ability to constrain cosmological models is limited by the systematics of relating the galaxy distribution to the underlying matter field.  For example, through a comparison of galaxy samples from 2dFGRS and SDSS, \citet{sanchez/cole:2007} find that differing scale-dependent biasing between red and blue galaxies cause the cosmological constraints derived from the samples to differ at the $\sim 2\sigma$ level.
These systematic problems are even more severe in the SDSS Luminous Red Galaxy (LRG) sample.  With its large effective volume \citep{tegmark/etal:2006}, this sample potentially provides the tightest constraints on the shape of the linear power spectrum.  However, because of the low number density of the LRGs and their high bias, the LRG power spectrum potentially requires a large nonlinear correction.
As a consequence, several authors have encountered problems when combining the LRG power spectrum results of \citet{tegmark/etal:2006} with CMB data sets.  \citet{dunkley/etal:2008} find that when combining with the 5 year WMAP results, the best fit $\Omega_m$ varies systematically with the maximum $k$ from the LRG $P(k)$ included in the analysis.  When \citet{verde/peiris:2008} reconstruct the primordial power spectrum with minimal assumptions about its shape, they find that the LRG sample has less statistical power than the SDSS MAIN or 2dFGRS, despite its larger effective volume.  Since the authors restrict themselves to $k \leq 0.1$, they are unable to constrain both the nonlinear correction and the primordial power spectrum.  The difficulties in both analyses stem from the degeneracy between the power spectrum shape and the potentially large nonlinear correction amplitude.   Fig.~\ref{fig:qnlplotnew} shows that the best fit \citet{tegmark/etal:2006} nonlinear correction is above the statistical error in each $k$-bin for $k \gtrsim 0.07$, while $k \lesssim 0.1 -0.15$ is typically used in cosmological parameter analyses.  Moreover, the $k$-dependence of this large nonlinear correction must be accurate to avoid introducing systematic errors.  The solution we propose in this paper is to first estimate a halo density field from the LRG density field.  The resulting field is nearly linearly biased with respect to the dark matter for $k \leq 0.2$.  The dashed curve in Fig.~\ref{fig:qnlplotnew} shows that the nonlinear correction between the dark matter and linear spectrum is much smaller than the correction applied in the \citet{tegmark/etal:2006} LRG analysis; it is also much easier to calibrate using $N$-body simulations.\\

The Halo Occupation Distribution (HOD) model \citep{seljak:2000,peacock/smith:2000,cooray/sheth:2002} allows us to describe the link between galaxies and dark matter, and provides a framework in which to analyze non-linearities contributing to the power spectrum.  Pairs of galaxies can be separated as one- or two-halo, i.e., occupying the same or distinct dark matter halos.  As discussed in \citet{seljak:2000} and \citet{schulz/white:2006}, scale dependence of the bias between galaxies and dark matter arises primarily because the one-halo galaxy term is enhanced relative to the dark matter more than the two-halo term.  To combat this effect, \citet{huff/etal:2007} introduce a configuration space band-power estimator which confines the one-halo contribution to $\sim 2-3$ $h^{-1}$ Mpc.  \citet{hamann/etal:2008} find that for measured galaxy power spectra (SDSS MAIN, 2dFGRS, and SDSS LRG samples), the nonlinearity can be sufficiently modeled as excess shot noise.\\

In extracting the underlying linear matter power spectrum from large galaxy redshift surveys, there are three main approaches to disentangling the effects of nonlinear evolution in the matter field, nonlinear redshift space distortions, and nonlinear galaxy biasing.  One can model the nonlinearities with a phenomenological fitting formula and marginalize over its free parameters when fitting cosmological parameters; this technique has been employed in analyses of SDSS LRGs \citep{tegmark/etal:2006} as well as in 2dFGRS \citep{cole/etal:2005}.  The work of \citet{yoo/etal:2008} instead starts with the HOD to predict the nonlinear modifications to large-scale clustering statistics.  In principal, cosmological and HOD parameters can be fit simultaneously.  In this forward approach, the measured large-scale clustering is compared with predicted clustering given the HOD parameters.  The method put forth in this work is an {\em inverse} method: we make use of phase information in the galaxy density field to estimate a halo density field.   The power spectrum of this field is nearly linearly biased with respect to the underlying nonlinear matter density field.  A future extension of this work is to also use the phase information to reconstruct the linear density field \citep{eisenstein/etal:2007}.  This combined approach would ``undo'' two sources of nonlinearity: galaxy bias and nonlinear dynamics.\\

In this paper we introduce our halo density field reconstruction method and use $N$-body simulations to test the method.  We first review previous approaches to the analysis of the LRG $P(k)$ in \S~\ref{backgroundsec}.  In \S~\ref{simsec} and \ref{mockcatalogs} we apply the CiC method of \citet{reid/spergel:2008} to produce high fidelity mock catalogs from 42 $N$-body simulations for three LRG redshift subsamples: NEAR, MID, and FAR.   \S~\ref{howtoreconstruct} describes our method for reconstructing a halo density field.  In \S~\ref{results} we present the nonlinear dark matter and mock catalog power spectra and covariance matrices.  We analyze in detail the effects of the Finger-of-God (FOG) compression algorithm employed in the \citet{tegmark/etal:2006} analysis of the LRG power spectrum.  We also present results for the angle-averaged redshift space spectrum, which is measured in \citet{percival/etal:2007}.  We demonstrate that the reconstructed halo density field is linearly biased with respect to the dark matter out to $k=0.2$ at the 2\% level for the MID and FAR subsamples and at the 4\% level for the NEAR sample.  For $k \leq 0.1$, the traditional regime used for cosmological parameter estimation, the discrepancy is $\leq 1\%$.  We also show that the beat-coupling model of \citet{hamilton/rimes/scoccimarro:2006} fits both the dark matter and mock catalog covariance matrices.  Finally we examine the structure of the redshift space distortions as a function of $k$.  Throughout this paper we adopt the cosmological parameters recommended from the latest WMAP5 analysis \citep{komatsu/etal:2008}: ($\Omega_m, \Omega_b, \Omega_{\Lambda}, n_s, \sigma_8, h)$ = (0.2792, 0.0462, 0.7208, 0.960, 0.817, 0.701).\\

\section{Background}
\label{backgroundsec}
\subsection{Summary of Previous Analyses of the SDSS LRG Power Spectrum}
Two power spectrum analyses were simultaneously published for the fifth data release of the SDSS.  \citet{percival/etal:2007} compute the monopole (i.e., angle-averaged) redshift space power spectrum of the combined sample of MAIN and LRG galaxies.  The method extends the FKP method \citep{feldman/kaiser/peacock:1994} to remove differential bias from the recovered spectrum, and optimally weights the galaxies according to their expected bias, which is determined from their luminosity.  This approach leads \citet{percival/etal:2007} to use a different selection function for the LRGs than adopted for our analysis.  We do not model the luminosities of the galaxies, and so precise evaluation of their method is not possible here.  However, we do compute the monopole redshift power spectrum of our mock catalogs.  This should compare most closely to the \citet{percival/etal:2007} method, though the weighting of the galaxies according to their luminosity will certainly alter the relative amplitudes of the one- and two-halo contributions to the power spectrum.\\

\citet{tegmark/etal:2006} focus their analysis entirely on the LRGs.  They examine three redshift subsamples: NEAR ($0.155 < z < 0.300$), MID ($0.300 < z < 0.380$), and FAR
($0.380 < z < 0.474$).  We model each of these samples separately in our analysis.  \citet{tegmark/etal:2006} use the PKL method \citep{tegmark/etal:2004:la} to estimate the real space galaxy-galaxy, galaxy-velocity, and velocity-velocity power spectra.  In linear theory the three spectra are related by \citep[Eqns. 2 and 3 in ] []{tegmark/etal:2006}:
\begin{eqnarray}
P_{gv}(k) & = & \beta r_{gv} P_{gg}(k) \label{pgv}\\
P_{vv}(k) & = & \beta^2 P_{gg}(k) \label{pvv}.
\end{eqnarray}
$\beta$ relates the amplitude of galaxy and velocity field fluctuations in linear theory (Eqn.~\ref{betaeqnch4}) and $r_{gv}$ is the dimensionless correlation coefficient between the velocity field and gravitational acceleration field.  The final estimate of the real space LRG power spectrum is a linear combination of all three, though it is dominated by the scaled redshift space monopole power spectrum.  Their method assumes that the parameters relating the galaxy and velocity power spectra, $\beta$ and $r_{gv}$, are scale independent, and they only use data with $k < 0.09 \; h \; {\rm Mpc}^{-1}$ to estimate these factors.  To make this assumption more accurate for the LRG density field, \citet{tegmark/etal:2006} compress FOGs before estimating the power spectrum.  The algorithm is detailed in \citet{tegmark/etal:2004:la}, but to our knowledge has not been extensively tested on accurate LRG mock catalogs.  The algorithm finds groups of LRGs by a Friends-of-Friends (FoF) algorithm \citep{frenk/etal:1988} and then isotropizes them by scaling all radial separations from the group center.  Two objects are considered friends when
\begin{equation}
\left[\left(\frac{r_{\parallel}}{10}\right)^2 + r_{\perp}^2 \right]^{1/2} \leq \left[ \frac{4}{3} \pi \bar{n}(1+\delta_c) + r_{\perp,max}^{-3}\right]^{-1/3}
\label{fogeqn} 
\end{equation}
with $\delta_c = 200$ and $r_{\perp,max} = 5  \; h^{-1}$ Mpc; the RHS of Eqn.~\ref{fogeqn} is 2.3 $h^{-1}$ Mpc for the SDSS LRG NEAR and MID subsample number densities, and 2.8 $h^{-1}$ Mpc for the FAR sample.  For comparison, the virial radius of a $10^{14} h^{-1} M_{\sun}$ halo is $\sim 1 \; h^{-1}$ Mpc ($\sim 2$ for a $10^{15} h^{-1} M_{\sun}$ halo).  Since the line-of-sight (LOS) distance is compressed in the LHS of Eqn.~\ref{fogeqn}, the algorithm will compress objects well beyond the virial radius of the typical host halo.  After examination of the NEAR, MID, and FAR subsamples, \citet{tegmark/etal:2006} conclude that the expected decrease in bias and increase in clustering amplitude with redshift cancel within the observational errors on the subsample spectra.  They therefore apply no corrections for evolution of the galaxy population with redshift, and compute a single power spectrum for the entire sample.\\

\subsection{Nonlinear Power Spectrum Models}
The $Q_{NL}$ model was introduced in \citet{cole/etal:2005} as a fitting formula relating the linear and galaxy power spectra:
\begin{equation}
\label{qnleqn}
P_{gal}(k) = \frac{1+Q_{NL} k^2}{1 + Ak} P_{lin}(k).
\end{equation}
Both from examining mock catalogs and the HOD model analytically, \citet{cole/etal:2005} fix $A = 1.4$ (redshift space) and $A = 1.7$ (real space) to model the `previrialization' suppression of power on large scales \citep{lokas/etal:1996}.  $Q_{NL}$ is considered a nuisance parameter whose amplitude is set both by nonlinearities in the matter field and by the scale-dependent bias of the galaxy sample.\\

\citet{eisenstein/seo/white:2007} examine in detail the damping of BAO features in the power spectrum through differential motion of pairs of tracers separated by the BAO scale.  
For our real space matter power spectrum we adopt their model
\begin{equation}
\label{Psmear}
P_{\rm smear}(k) = P_{lin}(k) e^{-k^2/2k_{BAO}^2} + P_{\rm no \; wiggles}(k) \left(1 - e^{-k^2/2k_{BAO}^2}\right).
\end{equation}
$P_{\rm no \; wiggles}(k)$, defined by Eqn. 29 of \citet{eisenstein/hu:1998}, is a smooth version of $P_{lin}(k)$ with the baryon oscillations removed.  This means the baryon oscillations remaining in $P_{smear}(k)$ are damped completely for $k \gg k_{BAO}$ and unaltered for $k \ll k_{BAO}$.  In \S~\ref{dmresults} we find $k_{BAO} = 0.14 \; h \; {\rm Mpc}^{-1}$ for our real space matter power spectrum, in agreement with the value reported in \S 6 of \citet{eisenstein/seo/white:2007}; they also find that Eqn.~\ref{Psmear} is a good approximation in redshift space, but with smaller $k_{BAO}$ than in real space.\\

\citet{percival/etal:2007} assume that both $P(k)$ and galaxy bias are linear, and show that the resulting fits to $\Omega_m$ for $ 0.01 < k < 0.06 \; h \; {\rm Mpc}^{-1}$ and $ 0.01 < k < 0.15 \; h \; {\rm Mpc}^{-1}$ are discrepant at the 2-3$\sigma$ level, demonstrating the need for nonlinear modeling to extract robust cosmological information from the observed galaxy power spectrum.  The LRG power spectrum measured in \citet{tegmark/etal:2006} requires a large nonlinear correction (see our Fig.~\ref{fig:qnlplotnew}).  They adopt the model
\begin{equation}
\label{tegnonlinear}
P_{LRG}(k) = b^2 P_{smear}(k, k_{BAO} = 0.1) \frac{1 + Q_{NL} k^2}{1+Ak}
\end{equation}
with $A$ fixed at 1.4 (M. Tegmark, private communication).  Their best fit value for the nuisance parameter is $Q_{NL} = 30$.  This model is only tested on power spectra of mock LRGs in {\em real} space and with no satellite LRGs.  We will compare this model to our more realistic mock catalogs in redshift space.\\

\subsection{Simplest HOD Model for the LRG Power Spectrum}
As we describe in more detail in \S~\ref{hodmodel}, the HOD model specifies $P(N_{LRG}|M)$, the probability that $N_{LRG}$ LRGs occupy a halo of mass $M$.  Following \citet{cooray/sheth:2002}, the nonlinear power spectrum can be separated into the contribution from pairs of galaxies occupying the same dark matter halo, $P^{1h}(k)$, and pairs occupying distinct dark matter halos, $P^{2h}(k)$.  In the large scale limit of Eqn.~128 in \citet{cooray/sheth:2002}, where the effect of finite size of halos can be safely neglected and the bias with respect to the dark matter can be approximated as linear, the LRG power spectrum is approximately
\begin{eqnarray}
P_{LRG}(k) & = & P^{1h}_{LRG}(k) + P^{2h}_{LRG}(k) \label{onetwopkeqn}\\
P^{1h}_{LRG} & = & \int dM \; n(M) \frac{\left< N_{LRG} (N_{LRG} - 1 ) | M) \right > }{\bar{n}^2_{LRG}} \label{onehaloterm}\\
P^{2h}_{LRG}(k) & = & b_{LRG}^2 P_{DM}(k)\label{twohaloterm}
\end{eqnarray}
where $n(M)$ is the halo mass function, $b_{LRG}$ is the large scale bias between the galaxies and matter, and $P_{DM}(k)$ is the nonlinear dark matter power spectrum.  This simple model assumes that the effects of satellite galaxies are simply to reweight halo pairs traced by central galaxies in $P^{2h}$, and to add a simple shot noise term given by Eqn.~\ref{onehaloterm}.  As we show below, this division of pairs works well in real space, and $P_{LRG}(k) - b_{LRG}^2 P_{DM}(k)$ is well described by $P^{1h}$.  However, pairs of galaxies in redshift space cannot be cleanly described by this separation.  For our HOD parameters, the halo velocity dispersion causes satellite galaxies to shift by $\left <\sigma^{2}\right>^{1/2} \sim 9$ $h^{-1}$ Mpc (with large tails) along the LOS.  This shifting causes power to be shuffled between scales and even the largest scale modes along the LOS to be damped by the FOG smearing; \citet{tinker:2007} provide a detailed treatment.  Our approach will be to circumvent many of the difficulties in redshift space by eliminating the FOGs due to satellite galaxies from the density field before computing $P_{LRG}(k)$.\\

\subsection{Modeling the Covariance Matrix}
The covariance matrix for a set of band powers $P(k_{i=1..N})$ is defined by
\begin{equation}
C_{ij} = \left<(P(k_i) - \bar{P}(k_i))(P(k_j) - \bar{P}(k_j)\right>.
\end{equation}
\citet{hamilton/rimes/scoccimarro:2006} showed that the largest nonlinear contribution to the covariance matrix is a beat-coupling term proportional to the power on the largest scale of the survey.  In any real survey of finite volume, the observed Fourier amplitudes are convolved with the survey window function $W_s$:
\begin{equation}
\delta_{obs}({\bf k}) = \int \delta({\bf k}') W_s({\bf k} - {\bf k}') \frac{d{\bf k}'}{(2\pi)^3} .
\end{equation}
The beat coupling contribution arises because neighboring Fourier modes $\delta({\bf k}) \delta(-{\bf k} - {\bf \epsilon})$ are coupled by nonlinear growth to the beat mode $\delta({\bf \epsilon})$.  When the DC mode of the survey is positive, all modes are amplified; when it is negative, all modes are suppressed.  This term can be large since the linear power spectrum drops so sharply with $k$.  \citet{hamilton/rimes/scoccimarro:2006} emphasize that this term does not contribute to the covariance of power measured from ensembles of traditional periodic box simulations, where the band power $k_i$ is averaged over $N_i$ complex modes by
\begin{equation}
\bar{P}(k_i) = \frac{1}{N_i} \sum_{j=1}^{N_i} \delta({\bf k}_{ij}) \delta^{\star}({\bf k}_{ij}).
\end{equation}
Here, the beat mode is ${\bf k}_{ij} - {\bf k}_{ij} = 0$, the DC mode.  However, since our simulations allow the DC mode to vary \citep{sirko:2005}, we will capture this term.  We therefore model our covariance matrix as the sum of the usual diagonal Gaussian and shot noise contributions and the beat-coupling contribution:
\begin{equation}
\label{cijmodel}
C_{ij,model} = \frac{1}{N_{i}}\left(P(k_i) + \frac{\alpha_{shot}}{\bar{n}}\right)^2 \delta^{K}_{ij} + 4\beta_{beat}R_{\alpha} P(k_i)P(k_j)\delta_{DC}^2(z).
\end{equation}
Here $\delta^{K}_{ij}$ is the Kronecker delta, and $\delta_{DC}^2(z) =  P_{L}(k=0,z=0)/V_{sim} \times D^2(z)/D^2(z=0)$ is the variance of the DC mode linearly evolved to redshift $z$ in the simulation volume $V_{sim} = L^3$ \citep{sirko:2005}.  In perturbation theory $R_{\alpha} \approx 2.62$ \citep{hamilton/rimes/scoccimarro:2006}.  We introduce $\alpha_{shot}$ and $\beta_{beat}$ to allow for excess shot noise and variation in the amplitude of the beat coupling term, though we expect both parameters to be $\sim 1$.  We show in \S~\ref{results} that Eqn.~\ref{cijmodel} provides a good model for both the dark matter and LRG covariance matrices.  For the dark matter, the shot noise contribution is negligible.\\

\citet{neyrinck/szapudi/rimes:2006} showed that Poisson fluctuations about the mean halo mass function introduce variance in the amplitude of the one-halo contribution to the dark matter power spectrum that can dominate the covariance matrix in the nonlinear regime.  Since our reconstruction of the halo density field seeks to eliminate the one-halo contribution from galaxies, we expect the covariance to be smaller for the reconstructed halo density field than for the original galaxy sample.\\

\subsection{Error Estimates on $P(k_i)$ and $C_{ij}$}
For the dark matter, errors on the bandpowers are straightforward.  They are simply the diagonal terms of the inverse covariance matrix divided by the number of simulations.  We use the model covariance matrix in Eqn.~\ref{cijmodel} to compute the inverse:
\begin{equation}
\sigma_{P(k_i)}^2 = \frac{1}{N_{sim}} ({\bf C}_{model}^{-1})_{ii}.
\end{equation}
Of course, the beat coupling leads to off-diagonal terms in both the covariance matrix and its inverse; these terms must be included when estimating errors on model parameters.\\

We also estimate an error on our estimate of the covariance matrix.  We ignore correlations between the $C_{ij}$'s in our error estimates, though they are certainly present.\\
\begin{equation}
\label{cijerrDM}
\sigma_{C_{ij}}^2 = \frac{1}{N_{sim}} \left<\left((P(k_{i}) - \bar{P}(k_{i}))(P(k_{j}) - \bar{P}(k_{j})\right)  - C_{ij})^2\right>.
\end{equation}

The situation for the mock catalogs is more tricky.  For each of the $s = 1,..,N_{sim}$ $N$-body simulations we produce $m = 1,..,N_{mocks}$ mock catalogs using our fixed HOD parameters to reduce the shot noise contribution.  We define
\begin{eqnarray}
C_{ij,HOD} & = & \left< (P_{sm}(k_i) - \bar{P}_s(k_i)) (P_{sm}(k_j) - \bar{P}_s(k_j)) \right> \label{cijHOD}\\
C_{ij,red} & = & \left<(\bar{P}_{s}(k_i) - \bar{P}(k_i)) (\bar{P}_{s}(k_j) - \bar{P}(k_j)) \right> \label{cijreduced}\\
C_{ij,tot} & = & \left< (P_{sm}(k_i) - \bar{P}(k_i)) (P_{sm}(k_j) - \bar{P}(k_j)) \right> \label{cijtot}\\
C_{ij,tot} & = & C_{ij,HOD} + C_{ij,red}. \label{cijsum}
\end{eqnarray}
Here $P_{sm}(k_i)$ denotes the band power for mock catalog $m$ populating simulation $s$, $\bar{P_s}(k_i)$ denotes the band power in a single simulation $s$ averaged over the $m$ mock catalogs populating $s$, and $\bar{P}(k_i)$ denotes a band power averaged over the entire set of $N_{mocks} \times N_{sim}$ catalogs.  $C_{ij,tot}$ is the covariance matrix for this set of mock catalogs.  $C_{ij,HOD}$ is the variance introduced by sampling the same matter density field with different mock catalog realizations and $C_{ij,red}$ is the reduced covariance of the power spectra from each simulation after averaging over $N_{mocks}$ catalogs in each simulation.  The expected error on bandpower $P(k_i)$ is then
\begin{eqnarray}
P(k_i) = \left< (P_{sm}(k_i) \right> \label{pkLRGavg}\\
\sigma^2_{P(k_i)} =  \frac{1}{N_{sim}} ({\bf C}_{model,red}^{-1})_{ii}. \label{pkLRGvar}
\end{eqnarray}
We again neglect the covariance of the covariance matrix elements to estimate the errors on $C_{ij,tot}$, $C_{ij,red}$, and $C_{ij,HOD}$ from their variance over the simulations.  These errors are only used for fitting the parameters $\alpha_{shot}$ and $\beta_{beat}$, so inaccuracies in the errors should make our estimates noisier, but not biased.\\

\section{Simulations}
\label{simsec}
We use the publicly-available Tree-Particle-Mesh (TPM) code \citep{bode/ostriker:2003} to run 42 periodic box simulations with the parameters selected in Chapter 3 of \citet{reid:phd}: $L_{box} = 558$ $h^{-1}$ Mpc and $N_{p} = 512^3$, corresponding to a particle mass $M_p = 1.43 \times 10^{11} M_{\sun}$.  This resolution ensures that there are at least 50 particles per halo populated with an LRG for our best fit HOD parameters.  We allow the slight change in the halo mass function in our lowest mass bin to be absorbed by a slight change in the best fit HOD parameters for central galaxies.  As outlined in Appendix A, we performed a battery of tests to ensure that the large and small scale clustering and velocity statistics were unaffected by our choice of mass resolution for the simulation.\\

Initial conditions for the simulations were generated with the publicly available {\it ic} code \citep{sirko:2005}.  We updated the code to use the mt19937 ``Mersenne Twister'' random number generator \citep{matsumoto/nishimura:1998} to select the initial density mode realizations.  {\it ic} was designed to generate initial conditions for an ensemble of periodic box simulations that will match real-space statistical properties such as the mass variance in spheres and $\xi(r)$.  For a periodic box simulation with comoving side length $L_{uni}$ at $z=\infty$, the initial modes are drawn from a convolved power spectrum $P_{L_{uni}}(k)$ for which $P_{L_{uni}}(0)/L_{uni}^3$ is the variance of the DC mode in a volume equal to the initial periodic box volume.  The DC mode is assumed to evolve linearly, and so its effects can be mimicked to first order in Lagrangian perturbation theory by a slight change in the cosmological parameters of the box as a function of the value of the DC mode in the realization (see Eqns. 19 - 22 of \citet{sirko:2005}).  The relation between the scale factor in the box cosmology, $a_{box}$, and the scale factor in the universe cosmology, $a_{uni}$, is
\begin{equation}
\label{aboxuni}
\frac{a_{box}}{a_{uni}} = 1 - \frac{1}{3} \frac{D_{uni}(a_{uni})}{D_{uni}(1)} \Delta_{o}
\end{equation}
where $\Delta_{o}$ is the amplitude of the DC mode at $a_{uni} = 1$, and the growth function $D_{uni}$ is evaluated using the universe cosmology.  In this scenario, each simulation in the ensemble represents an equal initial comoving volume, $L_{uni}^3$.  Once the universe has evolved to scale factor $a_{uni}$, the comoving size of the box is $L_{box} = L_{uni} (a_{box}/a_{uni})$.  Because overdense regions expand more slowly than underdense regions, we must weight the simulations by $(a_{box}/a_{uni})^3$ to obtain volume-averaged quantities.  This improvement over a set of simulations for which the DC mode is precisely 0 is crucial to the analysis in this paper, since we wish to extract accurate matter and halo power spectra as the density field enters the nonlinear regime, along with accurate estimates of band power covariances.\\ 

We output the dark matter particle positions and velocities at time steps nearest to universe redshifts $z_{NEAR} = 0.235$, $z_{MID} = 0.342$, and $z_{FAR} = 0.421$, the galaxy-weighted mean redshifts of the three redshift subsamples analyzed in \citet{tegmark/etal:2006}.  Eqn.~\ref{aboxuni} shows that for a snapshot of the universe at fixed redshift, each simulation should be output at a different $z_{box}$ that increases with $\Delta_{o}$.  Since the TPM code outputs particle data at the nearest whole time step to the desired redshift, the difference between the target and output redshift will vary slightly with $\Delta_{o}$.
We correct for this effect exactly in the linear regime by scaling the power spectra from each simulation by $(D(z_{target})/D(z_{output}))^2$ before computing averaged quantities.  Without this scaling, we get the same average dark matter power spectra; changes in the covariance estimates are well below 1\%.  Any nonlinear corrections to this scaling will have an even smaller effect, and so we safely neglect them.\\

\subsection{Calculating Power Spectra}
Because of the low number density of LRGs, the shot noise correction is large compared to the $P(k)$ of the continuous density field sampled by the LRGs.  We use an FFT with $N = 512^3$ points, and present the power spectrum out to $k = 0.4 \; h$ Mpc$^{-1}$ = $0.14 k_{N}$, where $k_{N} = \pi/\Delta_{grid}$ is the Nyquist frequency.  Comparison with an $N=1024^3$ grid showed agreement at the $\sim 10^{-6}$ level.  Following the work of \citet{hockney/eastwood:1981}, \citet{jing:2005}, and \citet{baugh/efstathiou:1994b}, mock LRGs are distributed on the FFT grid using the triangular-shaped cloud (TSC), and the power spectrum is estimated at each ${\bf k}$ by \citep{jing:2005}
\begin{eqnarray}
P({\bf k}) & = & \frac{|\delta_{LRG,FFT}({\bf k})|^2 - P_{shot,TSC}({\bf k})}{W_{TSC}^2({\bf k})} \label{Pest}\\
P_{shot,TSC}({\bf k}) & = & \frac{1}{\bar{n}} \Pi_{i} \left[1 - \sin^2\left(\frac{\pi k_i}{2k_{N}}\right) + \frac{2}{15}\sin^4\left(\frac{\pi k_i}{2k_{N}}\right)\right] \label{Pshot}\\
W_{TSC}({\bf k}) & = & \Pi_{i} \left[{\rm sinc}\left(\frac{\pi k_i}{2k_{N}}\right)\right]^3, \label{windowcorrect}
\end{eqnarray}
where $\delta_{LRG,FFT}({\bf k})$ is the FFT of the LRG overdensity field and $k_i$ are the Cartesian components of ${\bf k}$.  Eqn.~\ref{Pest} corrects for FFT aliasing in the limit that $P({\bf k}+{\bf k_{N}})$ is shot noise dominated, which is the case here.  Finally, we average the values of Eqn.~\ref{Pest} over k-bands with $\Delta k_{box} = 0.0113$.\\

\subsection{Halo Catalogs}
\begin{figure}
\includegraphics*[scale=0.75,angle=0]{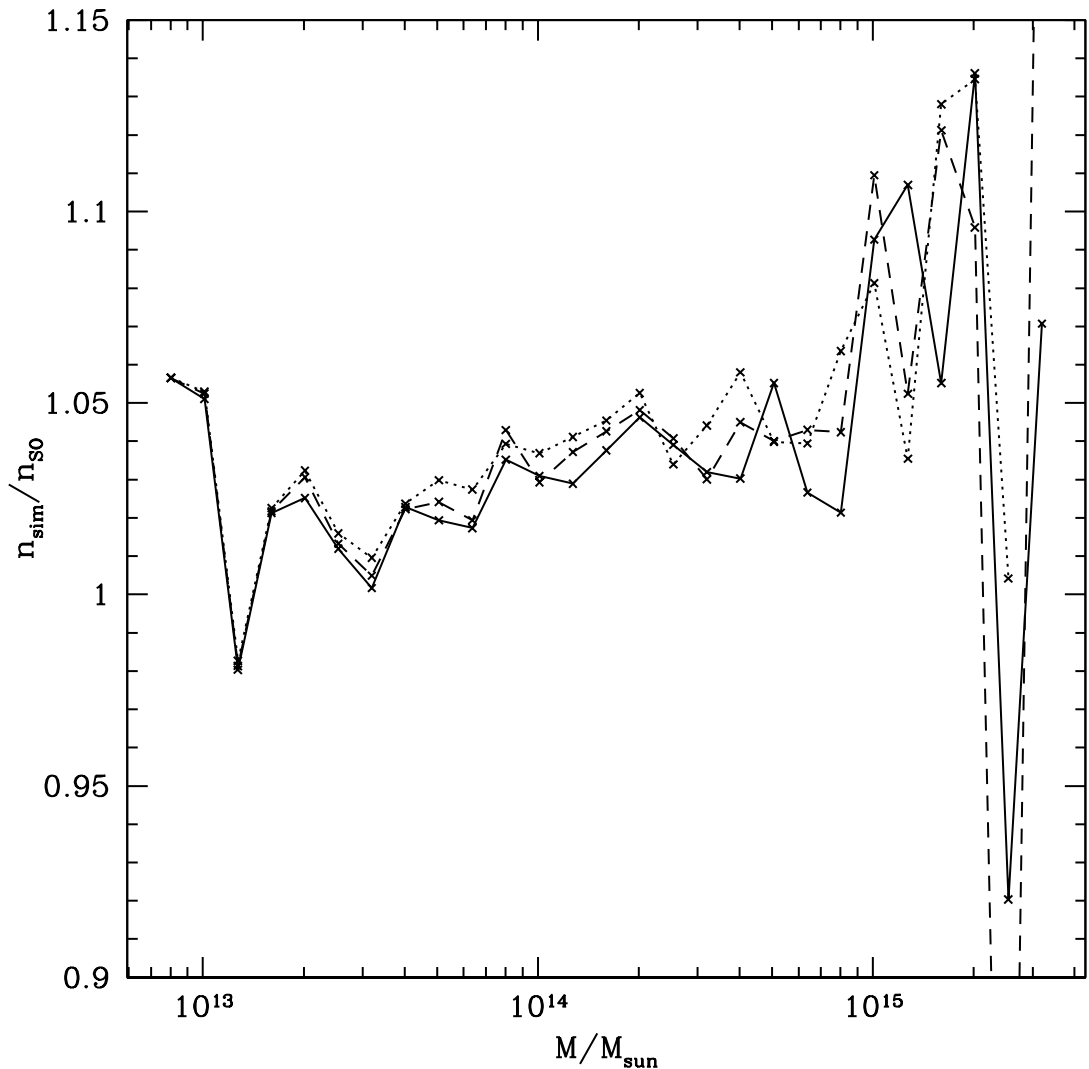}
\caption{\label{fig:MFratall} The ratio of the volume-weighted SO mass function from our simulations with the spline fit given in \citet{tinker/etal:2008} for the NEAR (solid line), MID (dashed line), and FAR (dotted line) in bins of $\Delta log_{10} M = 0.1$.  The agreement is within the stated accuracy of the \citet{tinker/etal:2008} spline fit ($\sim 5\%$).}
\end{figure}

We use the spherical overdensity (SO) halo finder code described in \citet{tinker/etal:2008} with $\Delta = 200 \rho_b$ to generate halo catalogs.  Fig.~\ref{fig:MFratall} shows the ratio of the volume-weighted SO mass function in our simulation set to the fitting function of \citet{tinker/etal:2008} down to $M = 7.15 \times 10^{12} M_{\sun}$.  There is good agreement at the level of accuracy claimed for their analytic fits ($\sim 5\%$).  The overabundance of halos in the lowest mass bins is likely due to the small number of particles per halo in those bins.  This slight modification of the mass function is unimportant for our purposes, since there are no satellite LRGs in these mass bins, and changes in the mass function are degenerate with changes to the probability of hosting a central galaxy as a function of halo mass, $N_{cen}(M)$.\\
\section{Mock Catalogs}
\label{mockcatalogs}
We use the technique described in \citet{reid/spergel:2008} to produce mock catalogs for the NEAR ($0.155 < z < 0.300$), MID ($0.300 < z < 0.380$), and FAR ($0.380 < z < 0.474$) redshift subsamples of \citet{tegmark/etal:2006}.  These samples contain roughly equal numbers of galaxies, and their power spectra have roughly the same amplitude on large scales.  We first review our HOD modeling assumptions, described in more detail in \citet{reid/spergel:2008}.\\
\subsection{\bf HOD Model}
\label{hodmodel}
The Halo Occupation Distribution (HOD) model assumes that the probability $P(N_{LRG}|M)$ of $N_{LRG}$ LRGs occupying a dark matter halo of mass $M$ at redshift $z$ depends only on the halo mass \citep[for a review, see ][] {cooray/sheth:2002}.  However, the application of the HOD formalism requires that we make several further assumptions about $P(N_{LRG}|M)$.  Detailed studies of dark matter halos and subhalos suggest a division of galaxies into central and satellite galaxies \citep{kravtsov/etal:2004}.
  The central galaxies are assumed to sit at the halo center, consistent with the observation that most ($\sim 80\%$) of the brightest cluster LRGs are found within $0.2r_{vir}$ of the center of the cluster potential well as traced by X-rays.  Satellite galaxies occur in the more massive halos already containing a central galaxy.  In high resolution simulations they can be directly associated with dark matter subhalos \citep{vale/ostriker:2006}; here we will assume they have the same distribution as the halo dark matter.  
We will use these functional forms with the five free parameters $M_{min}$, $\sigma_{log M}$, $M_{1}$, $M_{cut}$, and $\alpha$ to describe the mass dependence of the average halo occupation as a function of halo mass $M$:
\begin{eqnarray}
\left<N(M)\right> = \left<N_{cen}\right>(1 + \left<N_{sat}\right>) \label{censatsum}\\
\left<N_{cen}\right> = \frac{1}{2} \left[ 1 + {\rm erf} \left ( \frac{log_{10} M - log_{10} M_{min}}{\sigma_{log M}}\right)\right]\label{NcenM}\\
\left<N_{sat}\right> = \left(\frac{M - M_{cut}}{M_1}\right)^{\alpha} .\label{NsatM}
\end{eqnarray}
Halos are populated with a central galaxy with probability $N_{cen}(M)$.  Central galaxies are placed at the center of their host halos and assigned the peculiar velocity of their halos.  Halos with a central galaxy are populated with $N_{sat}$ galaxies, where $P(N_{sat}|N(M))$ is drawn from a Poisson distribution.  The position and velocity of a satellite galaxy is taken to be that of a randomly selected dark matter particle halo member.  We assign comoving redshift space position $s$ to an object in our mock catalogs using the conversion at $z_{box}$, the redshift at which the simulation data were output:
\begin{equation}
s = x_{LOS} + (1+z_{box})v_{p}/H(z_{box})
\end{equation}
where $x_{LOS}$ is the comoving distance along the line of sight in real space.\\
\subsection{Measurements of $N_{CiC}$ from SDSS and HOD parameter results}
The HOD model specified in \S~\ref{hodmodel} determines the number of groups with $n_{sat} = 0, 1, 2, ...$ satellite galaxies.  In \citet{reid/spergel:2008} we present the measurement of the SDSS Counts-in-Cylinders (CiC) group multiplicity function $N_{CiC}(n_{sat})$, calibrate its relation to the true group multiplicity function, and use a maximum likelihood analysis to derive the parameters of $N_{sat}(M)$ in Eqn.~\ref{NsatM}.  The parameters of $N_{cen}(M)$ in Eqn.~\ref{NcenM} are constrained by the observed LRG number density and large scale clustering amplitude.  To derive HOD parameters for the mock catalogs used in this paper, we follow the technique of \citet{reid/spergel:2008} exactly, but we must first measure $N_{CiC}(n_{sat})$ separately for our three redshift subsamples.\\

\begin{figure}
\includegraphics*[scale=0.75,angle=0]{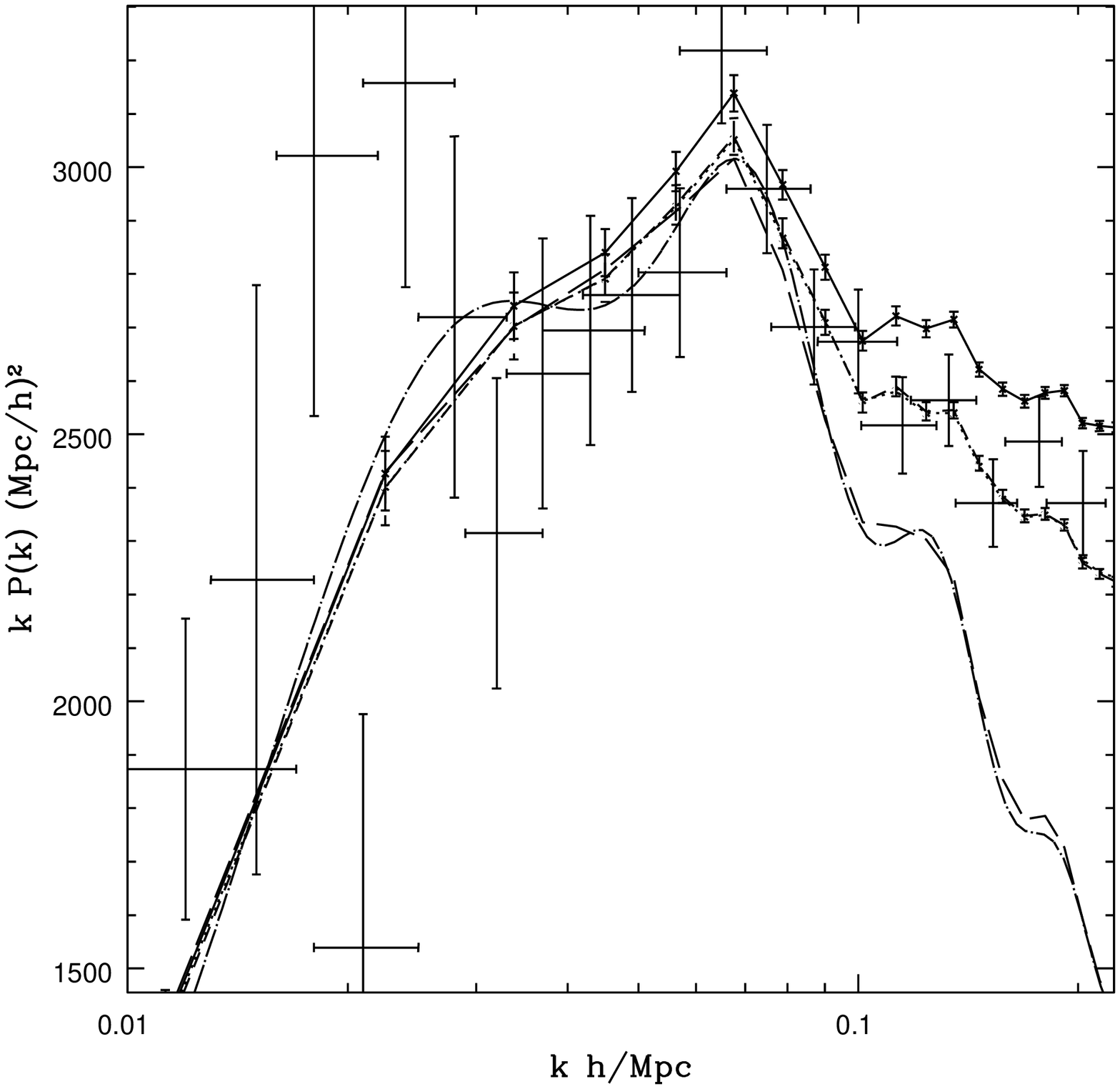}
\caption{\label{fig:pktegcomparech4}  $k P(k)$ for the NEAR (dotted curve), MID (dashed curve), and FAR (solid curve) mock galaxy samples with the FOG compression algorithm of \citet{tegmark/etal:2006} applied compared with their observed power spectrum (points with large error bars).  Our NEAR and MID samples are indistinguishable, so we show error bars only for the MID and FAR samples.  Our error bars are derived from the diagonal elements of the inverse $C_{ij,red}$ matrix.  We scale the real space $P(k)$ of \citet{tegmark/etal:2006} by $(0.8)^{-1}$ to approximate the redshift space monopole (i.e., angle averaged) power spectrum; this is the relation between the real and redshift monopole spectra if one neglects the small contributions from the quadrupole and hexadecapole detailed in their \S A3.  For comparison, the long dashed curve shows our simulation initial conditions drawn from the convolved power spectrum according to the {\em ic} algorithm, and the smooth dot-dashed curve shows the linear power spectrum for the cosmological parameters adopted in this study.}
\end{figure}

In Table~\ref{table:NEARMIDFARcic} we report our measurement of $N_{CiC}(n_{sat})$  from the SDSS DR4+ LRG sample for the NEAR, MID, and FAR redshift subsamples.  We introduce a few minor changes in this measurement from \citet{reid/spergel:2008}.  First, we include a boundary set of galaxies in the redshift direction, since FOGs may widely separate nearby pairs of LRGs; we verified that this has a small effect on the resulting CiC group multiplicity.  This buffer is $\Delta z = 0.007$ at the low redshift end of the sample and $\Delta z = 0.009$ at the high redshift; that is, equal to the maximum redshift separation for a CiC pair of galaxies at the sample boundary.  This causes the redshift ranges of the subsamples for which we measure $N_{CiC}(n_{sat})$ to be slightly different than the \citet{tegmark/etal:2006} NEAR, MID, and FAR samples.  Because the color cuts used to select the LRG sample produces a complex radial selection function, the effective number density of the two samples is slightly different.  As discussed in \citet{reid/spergel:2008}, the main difference between our sample number densities (Column 4 of Table~\ref{table:HODpartable}) and the result of integrating the \citet{zehavi/etal:2005a} model for the redshift dependence of $\bar{n}_{LRG}(z)$ (Column 5 of Table~\ref{table:HODpartable}) comes from our careful inclusion of objects from the imaging sample.  In the MID sample, this is an 8\% increase in the number density.  We have verified that once the imaging galaxies are accounted for along with the difference in redshift range of the NEAR and FAR samples, our simple number density estimate $n_{LRG} = N_{sample}/V_{sample}$ is in agreement with the expectations from the \citet{zehavi/etal:2005a} model.  Second, our redshift indicator used for objects without spectra must be altered, as the 4000 ${\rm \AA}$  break moves from the g to the r band at $z \simeq 0.4$ \citep[see Fig. 4 of][]{eisenstein/etal:2001}.  For our FAR sample we use the $r-i$ color as a redshift indicator.  We use the observed $N_{CiC}(n_{sat})$ for the NEAR, MID, and FAR subsamples to derive the subsample satellite HOD parameters $M_{cut}$, $M_{1}$, and $\alpha$ in Table~\ref{table:HODpartable} using the maximum likelihood technique presented in \citet{reid/spergel:2008}.  Since \citet{tegmark/etal:2006} find that the NEAR, MID, and FAR subsamples have consistent $P(k)$'s, we vary $\sigma_{log M}$ to match each redshift subsample to the large scale amplitude of the combined $P(k)$ reported in \citet{tegmark/etal:2006}.  $M_{min}$ is determined by $\bar{n}_{sim}$.  Fig.~\ref{fig:pktegcomparech4} shows the agreement of the large scale $P(k)$ with \citet{tegmark/etal:2006} when we apply their FOG compression algorithm to our mock catalogs.\\
\begin{deluxetable}{llll}
\tabletypesize{\scriptsize}
\tablewidth{0pt}
\tablecolumns{4}
\tablecaption{\label{table:NEARMIDFARcic} The final $N_{CiC}(n_{sat})$ group multiplicity function following the method in \citet{reid/spergel:2008} for our NEAR, MID, and FAR LRG subsamples.}
\tablehead{
\colhead{$n_{sat}$} & \colhead{$N_{\rm CiC,NEAR}(n)$} & \colhead{$N_{\rm CiC,MID}(n)$} & \colhead{$N_{\rm CiC,FAR}(n)$}}
\startdata
0 & 22921.71 & 24537.81 & 19109.71\\
1 & 1372.63 & 1301.29 & 664.28\\
2 & 170.01 & 153.40 & 61.94\\
3 & 41.85 & 25.59 & 6.08\\
4 & 15.16 & 9.07 & 2.04\\
5 & 2.11 & 2.04 & 1.00\\
6 & 1.01 & 1.01 & 0.00\\
7 & 1.02 & 0.00 & 0.00\\
8 & 0.03 & 0.00 & 0.00\\
\enddata
\end{deluxetable}
\begin{deluxetable}{lllllllllll}
\tabletypesize{\scriptsize}
\tablewidth{0pt}
\tablecolumns{11}
\tablecaption{\label{table:HODpartable} Mock catalog parameters.  Masses in units of $10^{14} M_{\sun}$ are for SO halos with $\Delta=200 \rho_b$; number densities are in units of $10^{-4} \, (h^{-1} \; {\rm Mpc})^{-3}$.  $z_{SDSS, \, CiC}$ is the redshift range of the SDSS LRG subsample and $\bar{z}_{sim}$ is the average simulation $z_{uni}$.  $\bar{n}_{sim}$ is the mock catalog number density while $\bar{n}_{model}$ is the expected number density obtained by integrating the \citet{zehavi/etal:2005a} model over the $z_{SDSS, \, CiC}$ range.  $P^{1h}$ is computed from Eqn.~\ref{onehaloterm} in units of ($h^{-1}$ Mpc)$^3$.}
\tablehead{
\colhead{Sample} & \colhead{$z_{SDSS, \, CiC}$} & \colhead{$\bar{z}_{sim}$} & \colhead{$\bar{n}_{sim}$} & \colhead{$\bar{n}_{model}$} & \colhead{$\sigma_{log M}$} & \colhead{$M_{min}$} & \colhead{$M_{cut}$} & \colhead{$M_{1}$} & \colhead{$\alpha$} & \colhead{$P^{1h}$}}
\startdata
NEAR & 0.162 - 0.300 & 0.2347 & 1.05 & 0.982 & 0.6 & 0.78 & 0.49 & 5.87 & 1.16 & 1759\\
MID & 0.300 - 0.380 & 0.3420 & 0.962 & 0.890 & 0.6 & 0.76 & 0.57 & 6.29 & 1.05 & 1480\\
FAR & 0.380 - 0.465 & 0.4216 & 0.470 & 0.422 & 0.9 & 2.27 & 1.38 & 5.97 & 0.78 & 2312\\
\enddata
\end{deluxetable}
\section{Reconstructing the Halo Density field}
\label{howtoreconstruct}
In the CiC technique detailed in \citet{reid/spergel:2008}, two galaxies are considered neighbors when their transverse comoving separation satisfies $\Delta r_{\perp} \leq 0.8$ $h^{-1}$ Mpc and their redshifts satisfy $\Delta z/(1+z) \leq \Delta v_p/c = 0.006$.  A cylinder should be a good approximation to the density contours of satellites surrounding central galaxies in redshift space, as long as the satellite velocity is uncorrelated with its distance from the halo center and the relative velocity dominates the separation of central and satellite objects along the line of sight.  Galaxies are then grouped with their neighbors by a FoF algorithm.  The reconstructed halo density field is defined by the superposition of the centers of mass of the CiC groups.  The CiC parameters were established as a balance between completeness and contamination in identifying pairs of galaxies at an early stage of this work based on FoF halo catalogs.  While the parameters and details of the method described here are sufficient for approximately recovering the halo density field power spectrum, the method could almost certainly be improved to more accurately recover groups of galaxies residing in the same dark matter halo.\\

\section{Results}
\label{results}
\subsection{Dark Matter}
\label{dmresults}
\subsubsection{Matter Power Spectrum}
\begin{figure}
\includegraphics*[scale=0.75,angle=0]{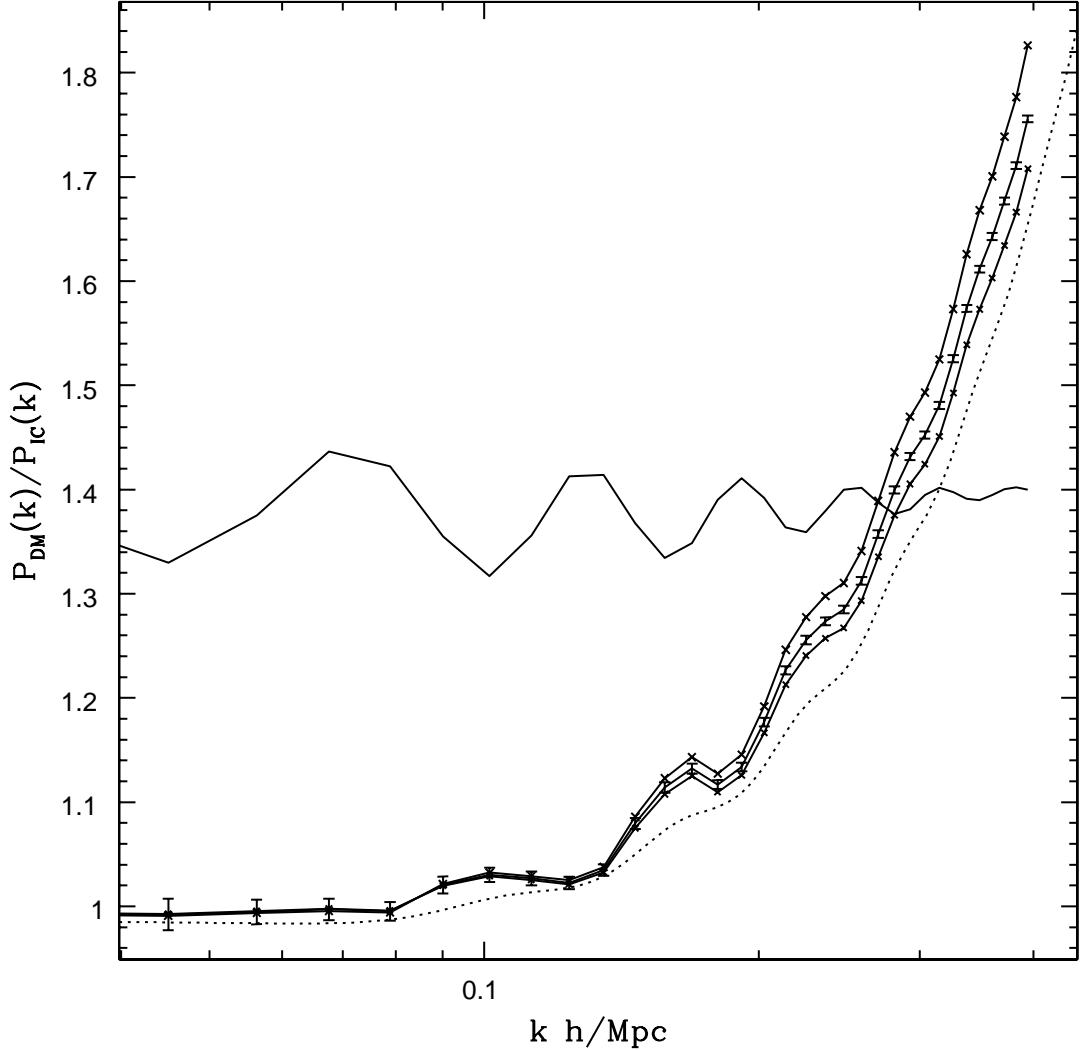}
\caption{\label{fig:DMnonlinear} The ratio of the volume-averaged matter power spectra at $z_{FAR} = 0.421$, $z_{MID}=0.342$, and $z_{NEAR}=0.235$ to the input power spectrum scaled by $D^2(z)/D^2(z=0)$; the amplitude of the nonlinear power spectrum increases as the redshift decreases.  The nonlinear correction to the matter power spectrum measured from our simulations is larger than expected from halofit \citep{smith/etal:2003}, shown as the dotted curve and evaluated at $z_{MID}$.  We also overlay $P_{IC}(k)/P_{\rm no \; wiggles}(k) + 0.4$ (solid curve oscillating about 1.4) to indicate the location of the baryon wiggles in the initial power spectrum.  The nonlinear evolution generates power that smoothly increases with $k$ and damps the baryon oscillations.  Error bars are shown only for the MID sample for clarity and are estimated from the inverse of the model covariance matrix defined in Eqn.~\ref{cijmodel}, where the shot noise is negligible and $\beta_{beat} = 1$; the beat-coupling term in Eqn.~\ref{cijmodel} correlates the bandpowers.}
\end{figure}

Fig.~\ref{fig:DMnonlinear} shows the ratio of the matter power spectrum to the power spectrum of the simulation initial conditions scaled by the expected linear growth $D^2(z)$ for our three redshift samples; the results at different redshifts are highly covariant since they are measured from the same set of simulations at relatively small separations in time.  Error bars are computed from the inverse of the model covariance matrix in Eqn.~\ref{cijmodel} with $\beta_{beat} = 1$ fixed.\\

As expected, the nonlinear correction grows as the redshift decreases.  The nonlinear evolution generates power that smoothly increases with $k$, and also damps the baryon oscillations.  The halofit nonlinear correction \citep{smith/etal:2003} (dotted curve, evaluated at $z_{MID}$) underestimates both the baryon wiggle suppression and the amplitude of the smooth increase in power; \citet{crocce/scoccimarro:2008} also find a disagreement between their $N$-body simulation results and the halofit nonlinear correction.\\

\begin{figure}
\includegraphics*[scale=0.75,angle=0]{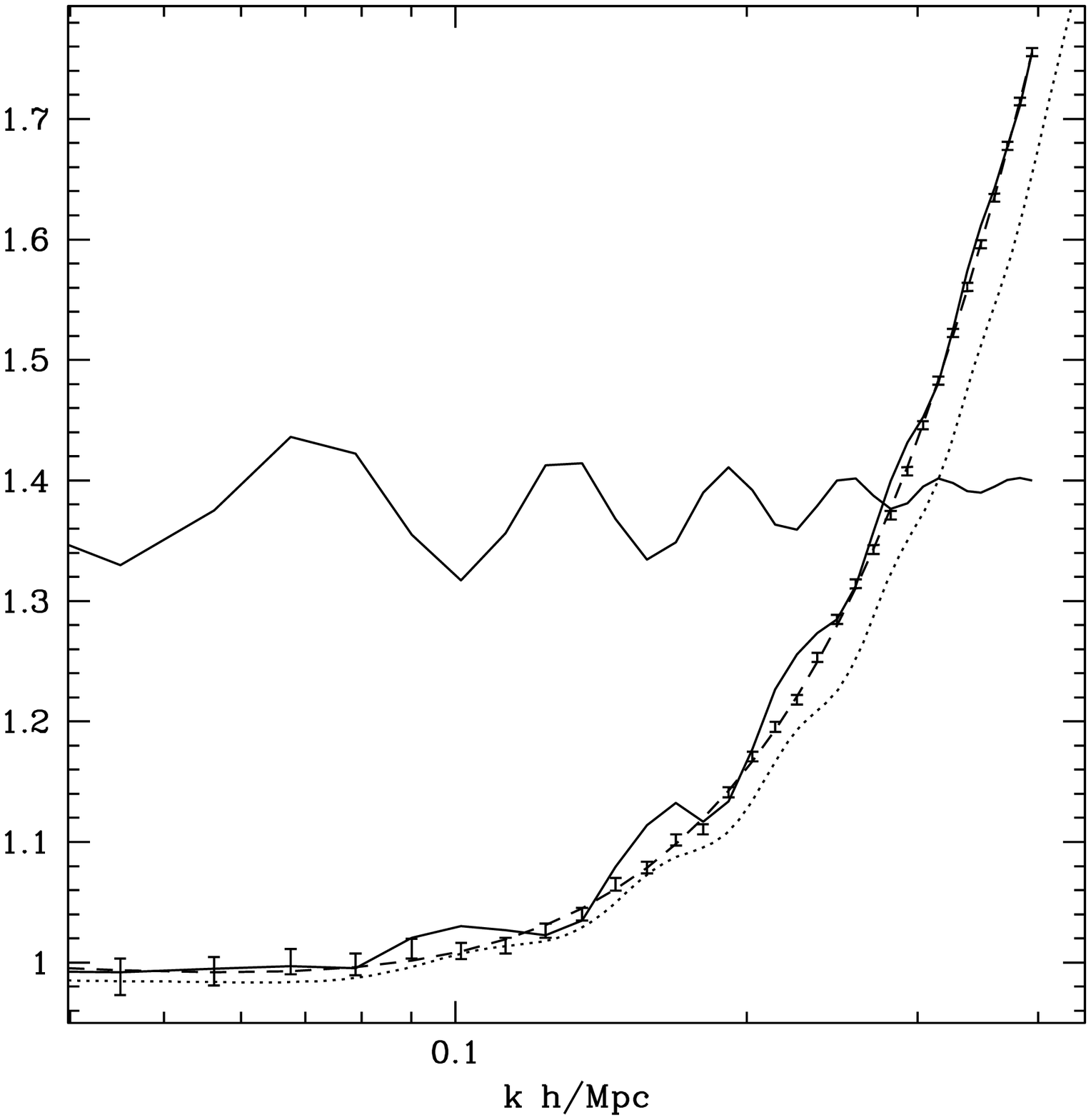}
\caption{\label{fig:DMnonlinear3} $P_{DM}(k)/P_{IC}(k)$ for the MID sample (solid curve, as in Fig.~\ref{fig:DMnonlinear}), $P_{DM}(k)/P_{\rm smear}(k)$ for $k_{BAO} = 0.14$ (points with error bars), and our polynomial fit $1.026 - 1.199 k + 11.06 k^2 - 8.426 k^3$ (dashed curve).  We also overlay $P_{IC}(k)/P_{\rm no \; wiggles}(k) + 0.4$ (solid curve oscillating about 1.4) to indicate the location of the baryon wiggles in the initial power spectrum.  Using error bars derived from the inverse of the model covariance matrix defined in Eqn.~\ref{cijmodel}, $\chi^2 = 21$ for 35 bandpowers and 5 parameters.  Comparison of the solid and dashed curves shows that $P_{\rm smear}(k)$ adequately models the BAO features, and that our third-order polynomial in $k$ is sufficient for $k \leq 0.4 \; h \; {\rm Mpc}^{-1}$.  The halofit \citep{smith/etal:2003} correction (dotted curve) shows better agreement with the smooth portion of the nonlinear correction, though halofit underestimates the nonlinear power by $4\%$ at $k=0.2 \; h \; {\rm Mpc}^{-1}$.} 
\end{figure}

Our fitting function to the nonlinear matter power spectrum $P_{DM}(k)$ aims to capture both the smearing of the acoustic peaks and a smooth increase in power with $k$.  We follow \citet{eisenstein/seo/white:2007} in defining $P_{smear}(k)$ (Eqn.~\ref{Psmear}), but adopt a slightly different fitting function for the smooth correction.  Our model is
\begin{equation}
\label{Pnlmat}
P_{DM}(k) = P_{\rm smear}(k;k_{BAO}) \left(a_0 + a_1 k + a_2 k^2 + a_3 k^3\right).
\end{equation}
For the MID sample with 35 bandpowers and 5 parameters, $\chi^2 = 21$.  We find $k_{BAO} = 0.14$, in good agreement with the values reported in \citet{eisenstein/seo/white:2007}.  In Fig.~\ref{fig:DMnonlinear3} we show that the $P_{\rm smear}(k)$ term accounts for the baryonic features in $P_{DM}(k)$, and the polynomial in $k$ adequately fits the smooth correction for the MID sample, while halofit underestimates the smooth correction by $\sim 4\%$ at $k=0.2$.  The NEAR and FAR fits are similar.  In Table~\ref{table:nonlinearfits} we list fits for the NEAR, MID, and FAR power spectra out to a maximum $k$ of 0.2 and 0.4 $h \; {\rm Mpc}^{-1}$.  When $k_{max,fit} = 0.2$ $h \; {\rm Mpc}^{-1}$, only the first three terms in the polynomial expansion are necessary for a good fit.\\

\subsubsection{Covariance Matrix}
\begin{figure}
\includegraphics*[scale=0.75,angle=0]{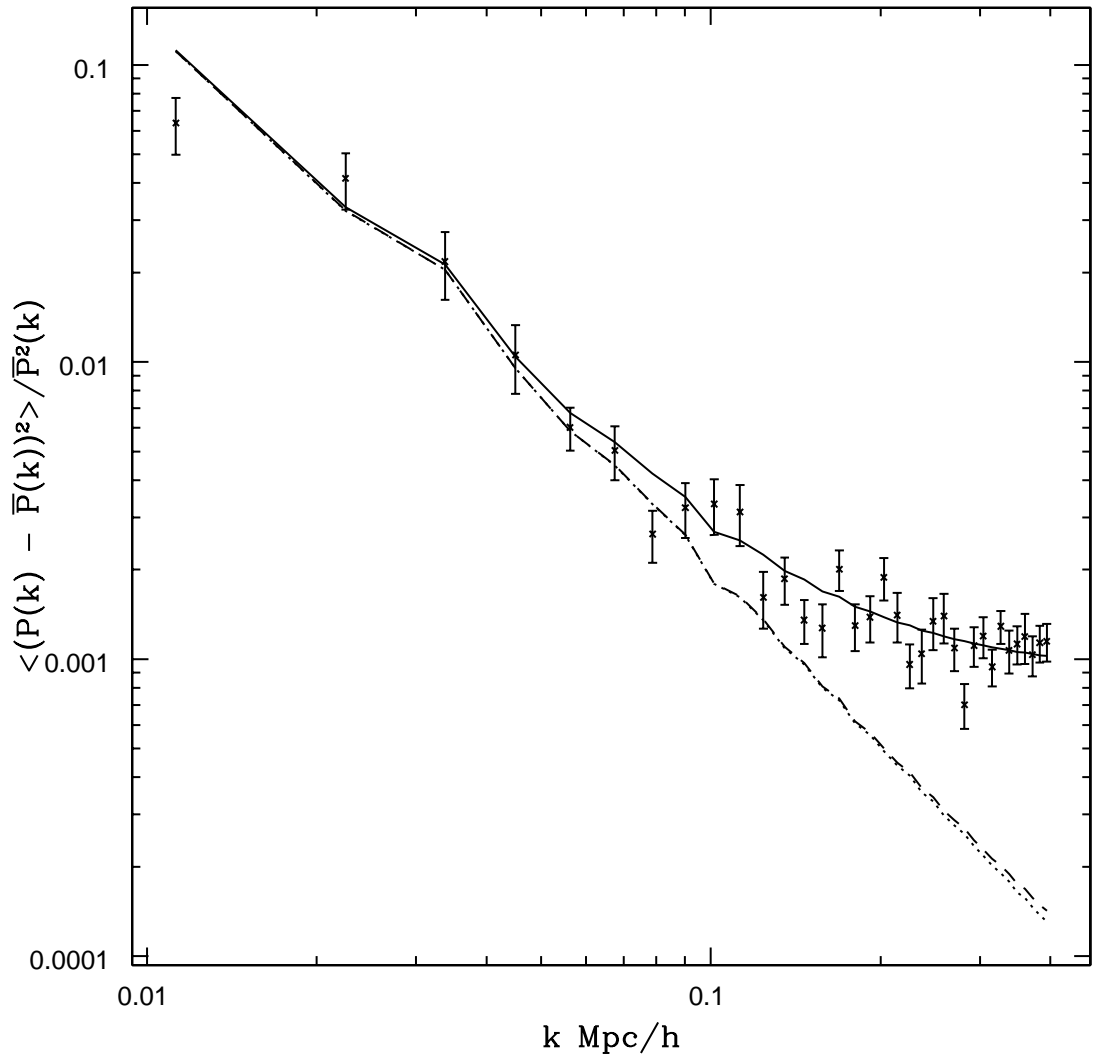}
\caption{\label{fig:diagcov} The points with error bars show the measured diagonal elements of the covariance matrix.  Error bars are estimated from the variance of $C_{ii}$ across our simulations.  The dotted curve shows the Gaussian prediction for the diagonal covariance, the inverse of the number of complex modes in each linearly-spaced $k$-bin.  The solid curve shows the sum of the Gaussian term and the beat-coupling term in \citet{hamilton/rimes/scoccimarro:2006}, which provides a good model for the full covariance matrix.  The dashed curve shows the inverse of the diagonal elements of the inverse covariance matrix.  We find that while the errors on the bandpowers are very close to the Gaussian expectation, the beat-coupling terms in the covariance matrix mean the bandpowers are correlated.} 
\end{figure}

%$\left<\left(P(k)_{i} - \bar{P}(k_i)\right) \left(P(k)_{j} - \bar{P}(k_j)\right)\right>/\bar{P}(k_i) \bar{P}(k_j)$
Fig.~\ref{fig:diagcov} shows diagonal elements of the normalized covariance matrix, $C_{ii}/\bar{P}(k_i)^2$.  We estimate the errors from the diagonal variances (Eqn.~\ref{cijerrDM}); these may not capture the true errors since off-diagonal elements should be present in the 8-point function as well.  Nevertheless, when we use these error estimates to compute $\chi^2$ for the model in Eqn.~\ref{cijmodel}, we find $\chi^2 = 1600$ for 1296 degrees of freedom ($0 \leq k \leq 0.4$); if we restrict the covariance matrix to the 196 elements with both $k$ bands between 0.056 and 0.21, we find $\chi^2 = 262$.  In this case there are no free parameters and we deem the model a reasonable fit.  If we allow the amplitude of the beat coupling term to vary, we find a best fit value 0.96 for the full matrix ($\chi^2 = 1564$) and 0.84 for the $0.056 \leq k \leq 0.21$ subsample ($\chi^2 = 218$).  We note that for the DC mode realizations of our 42 simulations, the variance is a factor of 0.83 lower than the expected variance (in agreement with the expected random variation for a single mode, $N_{sim}^{-1/2} = 15\%$).  We conclude that Eqn.~\ref{cijmodel} is an excellent fit to our dark matter covariance matrix.\\

Another point of interest in Fig.~\ref{fig:diagcov} is that the inverse of the diagonal elements of the inverse covariance matrix are nearly equal to the Gaussian expectation (dashed curve).  This means that while the beat-coupling does not introduce additional errors on the measurement of the bandpowers, the measured values will be covariant.  However, for a beat-coupling term in the form of Eqn.~\ref{cijmodel}, only information on the overall amplitude of $P(k)$ is lost.  However, large scale structure analyses traditionally marginalize over the amplitude of $P(k)$, since $\sigma_8$ and bias are degenerate.\\

\subsection{Mock Catalogs}
\subsubsection{Power Spectra}
Fig.~\ref{fig:pktegcomparech4} shows the agreement between our NEAR, MID, and FAR power spectra and the measurements presented in \citet{tegmark/etal:2006}.  We used this comparison to set by eye the large scale bias of our mock catalogs through $\sigma_{\log M}$, the width of $N_{cen}(M)$ in Eqn.~\ref{NcenM}.  The small discrepancy between the normalizations of the FAR and other catalogs at small $k$ could be eliminated with a slight variation in $\sigma_{\log M,FAR}$.  However, the FAR sample clearly has a different shape, and is consistent with trends in Fig.~6 of \citet{tegmark/etal:2006}, though in the SDSS sample the significance of these trends is less clear.  Our FAR sample has $\sim 3\%$ more power at $k=0.09$ and $\sim 10\%$ more power at $k=0.2$ than the NEAR and MID samples.\\

\begin{figure}
\includegraphics*[scale=0.75,angle=0]{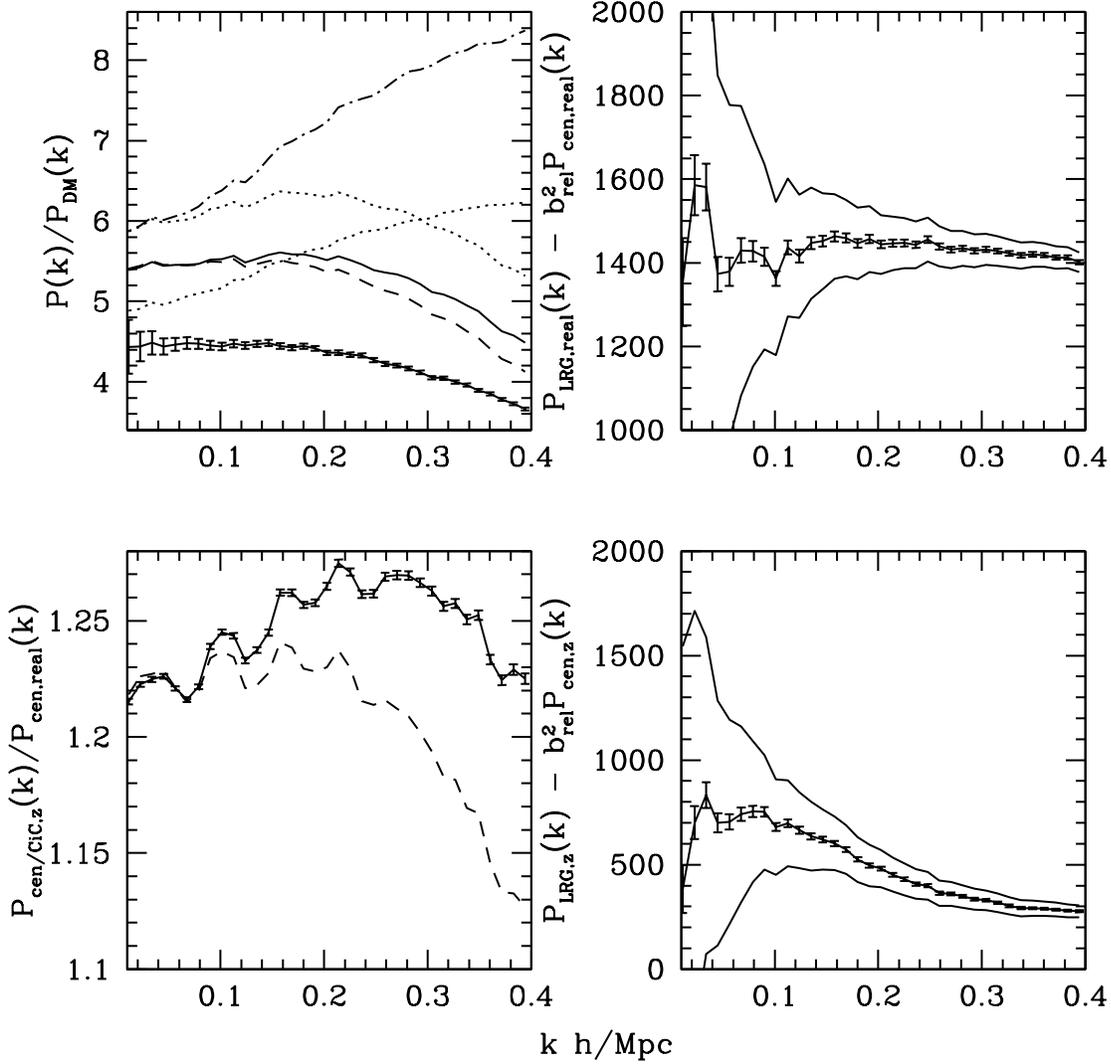}
\caption{\label{fig:LRGratMIDall} {\it Upper left}: The ratio of the monopole (i.e., angle-averaged) LRG $P(k)$ to the real space nonlinear dark matter spectrum for several MID subsamples: central LRGs only (solid curves); all LRGs (dotted curves); all LRGs after undergoing \citet{tegmark/etal:2006} FOG compression (dash-dot curve); and CiC groups (dashed curve).  For the first two subsamples we also show the real space $P(k)$, which have lower amplitude as $k \rightarrow 0$.  {\it Upper right}: The real space $P(k)$ for the central and satellite LRGs minus $b^2_{rel}$ times $P_{cen}(k)$, the power spectrum of the central LRGs in real space, for $b_{rel} = 1.037, 1.042, 1.047$.  {\it Lower left}: The solid curve shows the ratio of the central LRGs $P(k)$ in redshift space to the central LRGs $P(k)$ in real space, while the dashed curve shows the ratio for $P_{CiC}(k)$.  {\it Lower right}:  The redshift space monopole $P(k)$ for the central and satellite LRGs minus $b^2_{rel}$ times $P_{cen}(k)$, the power spectrum of the central LRGs in redshift space, for $b_{rel} = 1.037, 1.042, 1.047$.  Error bars are similar for all curves in a figure, and so we show only one set for clarity.}
\end{figure}

\begin{figure}
\includegraphics*[scale=0.75,angle=0]{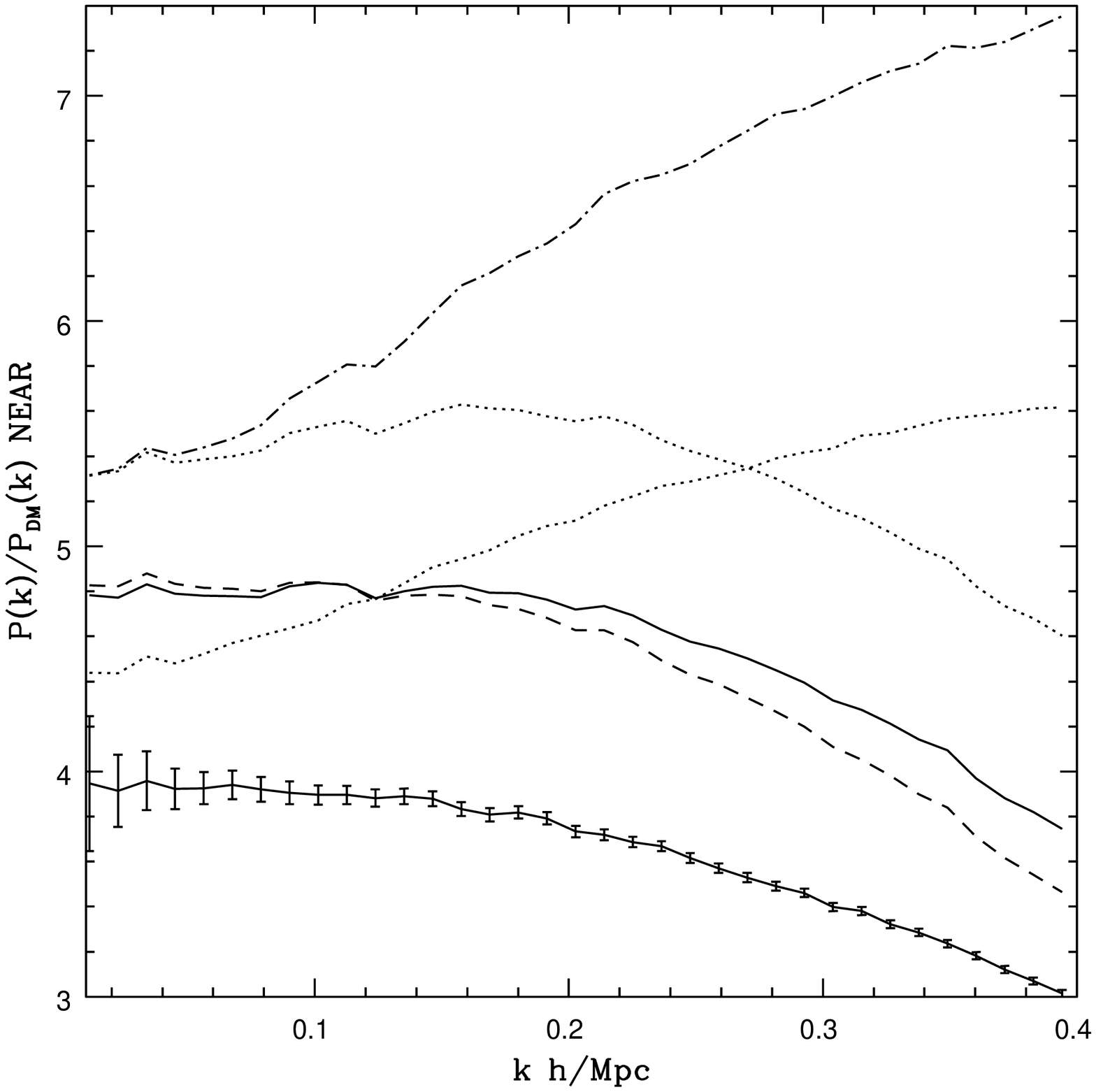}
\caption{\label{fig:LRGratNEAR} Same as Fig.~\ref{fig:LRGratMIDall} upper left panel, but for the NEAR sample.}
\end{figure}

\begin{figure}
\includegraphics*[scale=0.75,angle=0]{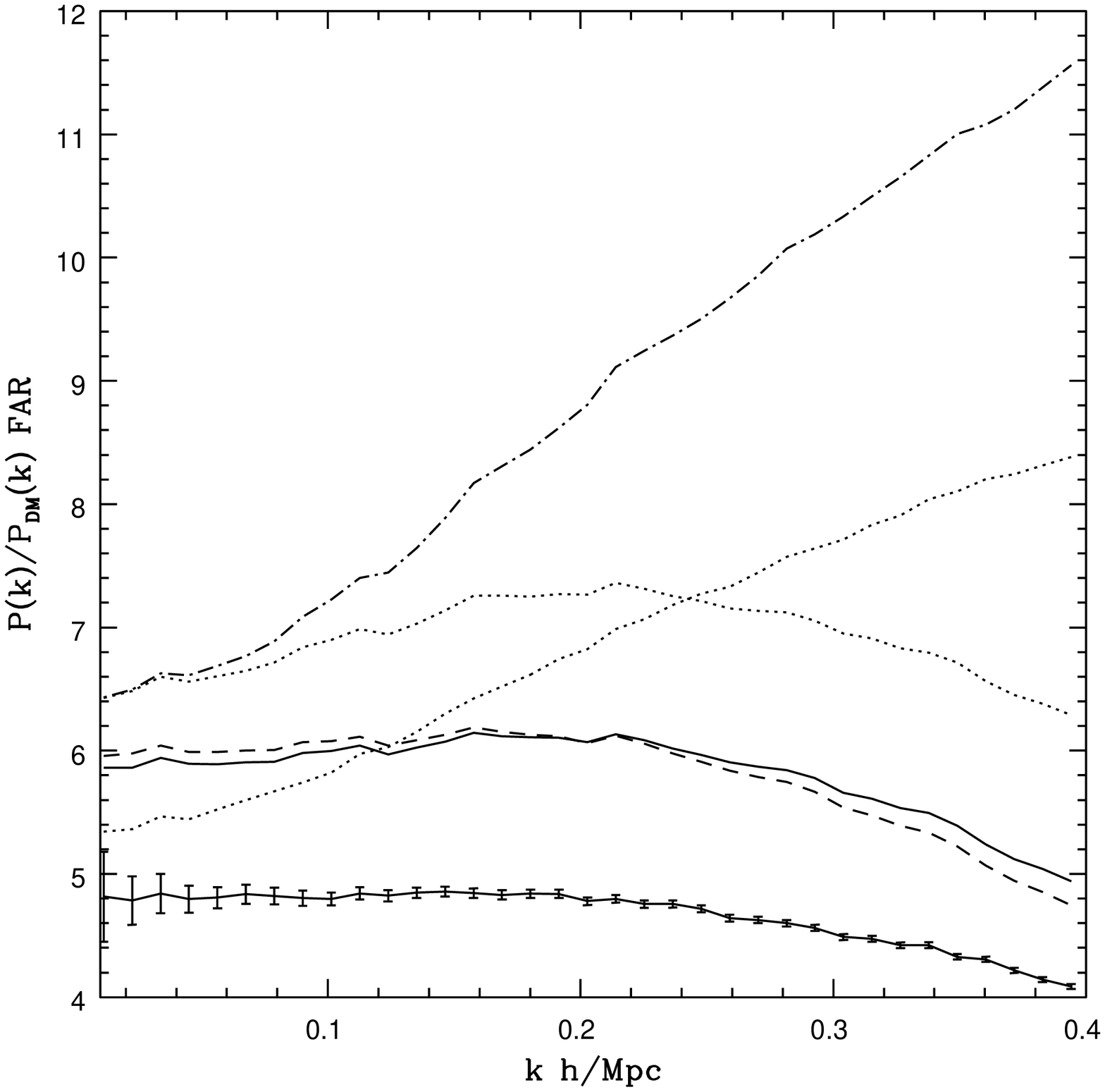}
\caption{\label{fig:LRGratFAR} Same as Fig.~\ref{fig:LRGratMIDall} upper left panel, but for the FAR sample.}
\end{figure}

Having established the agreement between our mock catalogs and the \citet{tegmark/etal:2006} observed $P_{LRG}(k)$, we now attempt to isolate the sources of nonlinearity in our catalogs and demonstrate that the power spectrum of the reconstructed halo density field is the best tracer of the underlying linear spectrum.  In Fig.~\ref{fig:LRGratMIDall} we analyze in detail the MID sample, while Figs.~\ref{fig:LRGratNEAR} and \ref{fig:LRGratFAR} show that the NEAR and FAR sample behave similarly.  The first major result is that there is no detectable deviation from a constant bias between the central LRGs and the dark matter in real space for $k \leq 0.1$, and the deviation at $k=0.2$ is $\leq 2\%$ for the MID and FAR subsamples; the discrepancy for the NEAR sample is $\sim 1\%$ at $k=0.1$ and 5\% at $k=0.2$ (solid curves with the lowest amplitude as $k \rightarrow 0$).  The lower left panel of Fig.~\ref{fig:LRGratMIDall} demonstrates that the main effect of using redshift space coordinates for the central LRGs is the further damping of the BAO signatures.  There is also a nonmonotonic but smooth variation that is $\leq 4\%$ between $k=0$ and $k=0.4$.  In the $k \leq 0.2 \; h$ Mpc$^{-1}$ regime, the direction of the deviation is opposite of that between the real space central LRGs and underlying dark matter, so that the redshift space central LRG power spectrum is nearly linearly related to the real space matter power spectrum at $k \leq 0.2 h \; {\rm Mpc}^{-1}$.\\

The inclusion of satellite LRGs in real space has two effects.  The linear bias is increased because the satellites upweight only the most massive, more highly biased halos traced by the LRGs.  Secondly, the satellite galaxies add a shot noise given by Eqn.~\ref{onehaloterm}.  The upper right panel of Fig.~\ref{fig:LRGratMIDall} shows that the difference between the $P(k)$ including satellites and $P(k)$ of the central LRGs only is well described by these two effects, with a relative bias $b_{rel} = 1.042$ and $P^{1h} \approx 1460$ ($h^{-1}$ Mpc)$^3$.  This latter value is in good agreement with 1480 ($h^{-1}$ Mpc)$^3$, the value computed directly from Eqn.~\ref{onehaloterm} using our input HOD parameters.  As $k \rightarrow 0.4$, the amplitude of the difference slightly diminishes due to the width of the satellite LRG density profile within the halos.  $P^{1h}$ can be $\sim 7-11\%$ of $P_{LRG}(k)$ at $k=0.1$, and $\sim 15 - 23\%$ at $k=0.2$ (see Table~\ref{table:HODpartable}).  If uncorrected for, this will be the dominant source of nonlinearity for LRGs owing to their low number density, which appears in the denominator of Eqn.~\ref{onehaloterm}.  Moreover, since the expected $P^{1h}$ varies across the NEAR, MID, and FAR samples, the power spectrum shape will necessarily vary with redshift, as is evident in Fig.~\ref{fig:pktegcomparech4}.\\

\begin{figure}
\includegraphics*[scale=0.75,angle=0]{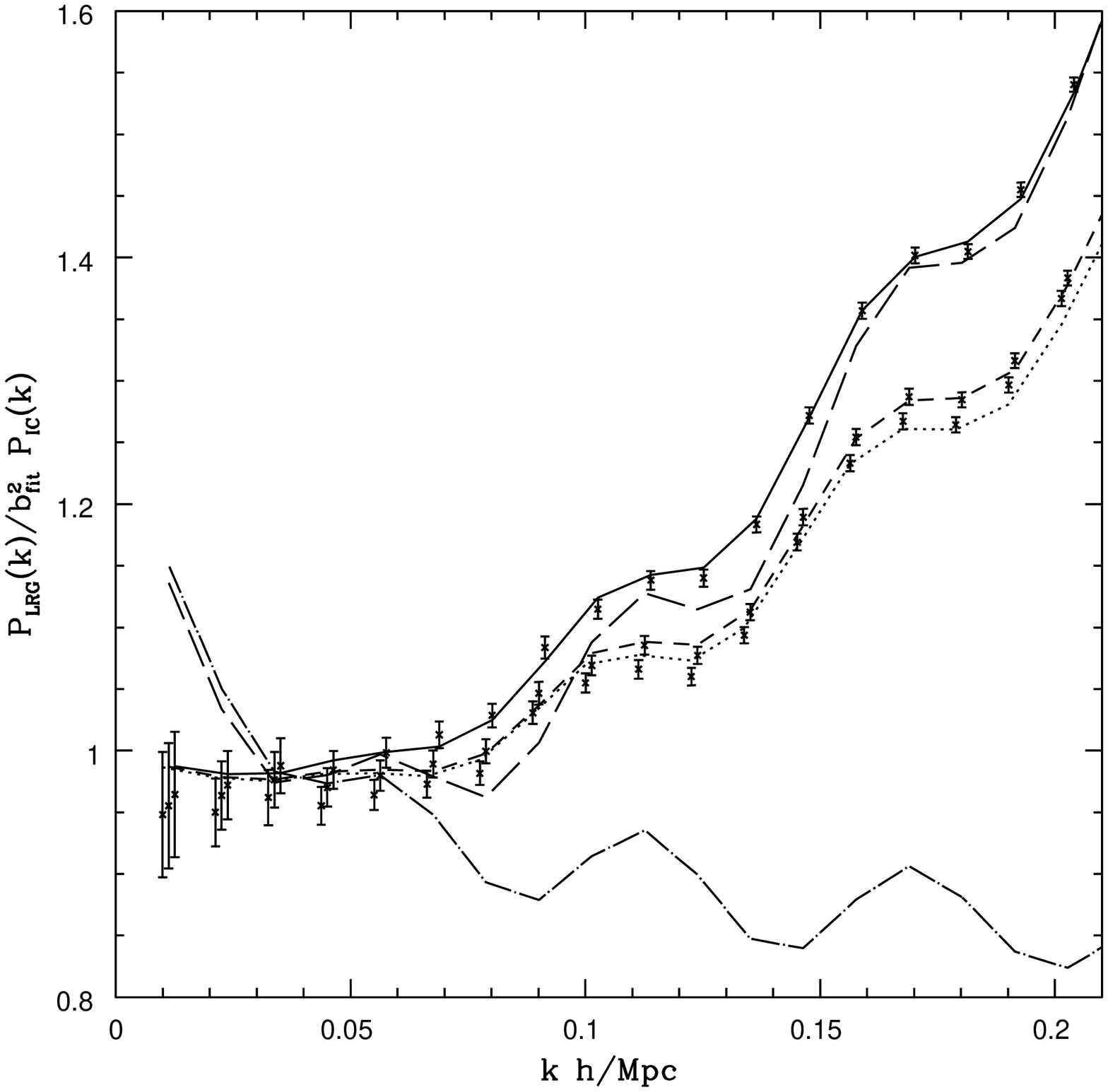}
\caption{\label{fig:qnlfitall}  We separately fit our NEAR, MID, and FAR FOG-compressed samples to the nonlinear correction model of \citet{tegmark/etal:2006} in Eqn.~\ref{tegnonlinear}.  Best fit parameters are $b^2_{NEAR} = 5.61$, $Q_{NL,NEAR} = 17.4$ (dotted curve); $b^2_{MID} = 6.15$ , $Q_{NL,MID} = 18.4$ (dashed curve); and $b^2_{FAR} = 6.67$, $Q_{NL,FAR} = 23.3$ (solid curve).  Each subsample is scaled by the input power spectrum and the best fit $b^2$ value to accentuate the variation in the power spectrum shapes for these three samples.  The NEAR and FAR data points and model curves have been shifted by -/+ 0.0013 $h \; {\rm Mpc}^{-1}$ for clarity.  The best fit $Q_{NL}$ vanilla model in \citet{tegmark/etal:2006}, $A = 1.4$ and $Q_{NL} = 31$ in Eqn.~\ref{qnleqn}, is shown by the long dashed curve.  We have adjusted the normalization to agree at $k \sim 0.05$.  The dot-dashed curve shows the ratio of the best fit linear power spectrum of \citet{tegmark/etal:2006} and the linear power spectrum adopted in this work, demonstrating that $Q_{NL}$ is degenerate with changes in the spectral shape, so that cosmological parameter information is lost to $Q_{NL}$.}
\end{figure}

We now consider the implications of the FOG-compression algorithm of \citet{tegmark/etal:2006}, which was designed to recover the real space $P_{LRG}(k)$.  Figs.~\ref{fig:LRGratMIDall} through \ref{fig:LRGratFAR} demonstrate that their compression algorithm (Eqn.~\ref{fogeqn}) is overly aggressive and adds more one-halo power than is present in the real space mock catalog (dash-dot curves).  This is the main source of the large nonlinear correction required in \citet{tegmark/etal:2006}.  However, Fig.~\ref{fig:qnlfitall} shows that the nonlinear correction model of \citet{tegmark/etal:2006} in Eqn.~\ref{tegnonlinear} accurately describes their resulting power spectrum for each of the redshift subsamples, so that the resulting cosmological constraints should not be biased.  The best fit $\Lambda$CDM - $Q_{NL}$ model in \citet{tegmark/etal:2006} (long-dashed curve of Fig.~\ref{fig:qnlfitall}) is consistent with being a weighted average over the NEAR, MID, and FAR subsample mock catalog power spectra.  The dot-dashed curve in Fig.~\ref{fig:qnlfitall} shows the ratio of the linear $P(k)$ for their best fit cosmological parameters with $P_{IC}(k)$ from our simulations.  These two models differ at a level larger than the statistical error bars shown in Fig.~\ref{fig:qnlplotnew}, even if we restrict the analysis to $k < 0.1$, and certainly if we can use the measurements between $k=0.1$ and $k=0.2$.  However, the $Q_{NL}$ parameter is degenerate with this difference in power spectra, so that with this nonlinear correction model we cannot distinguish the two linear spectra.\\

In redshift space the effect of the satellite galaxies on the angle-averaged power spectrum is more complicated.  At high $k$ the ratio $P(k)/P_{DM}(k)$ turns over as the suppression of power by the FOGs becomes more important than the addition of one-halo power.  In the bottom right panel of Fig.~\ref{fig:LRGratMIDall} we see that the difference between the all LRG and central LRG power spectra is no longer well-approximated by the constant $P^{1h} \sim 1480$ ($h^{-1}$ Mpc)$^3$ from Eqn.~\ref{onehaloterm}.  The one-halo pair separation in redshift space is large ($\sim 9$ $h^{-1}$ Mpc but with broad tails), so that the neat division between one-halo and two-halo pairs in real space no longer holds, and power will be transferred between $k$-bands in a complicated way.  The reconstructed halo density field method removes the satellite galaxies before computing $P(k)$, which allows us to avoid the necessarily complicated FOG modeling \citep{tinker:2007}.\\

The power spectrum of the reconstructed halo density field is very similar to the redshift space power spectrum of the central LRGs for the NEAR, MID, and FAR subsamples, as one would expect if the reconstruction is sufficiently accurate.  The slight difference is evident in the bottom left panel of Fig.~\ref{fig:LRGratMIDall}, and may be because our method puts the reconstructed halo at the center of mass of the CiC group, which will leave residual FOG smearing.  An improved method may be to put the reconstructed halo at the position of the brightest LRG, which should be closest to the halo center and have a smaller velocity with respect to its host halo.
For $k \leq 0.1$ the reconstructed halo density field does not detectably deviate from the shape of the underlying matter power spectrum for the NEAR and MID samples; for the FAR sample, the deviation is $\sim 1\%$.  In contrast, the FOG-compressed sample deviates from a constant bias at the 6\%, 7\%, and 10\% level between $k=0.05$ and $k=0.1$.  The reconstructed halo density field has only small deviations from a scale-independent bias out to $k=0.2$: the deviation is 4\%, 2.8\%, and 2.5\% for the NEAR, MID, and FAR sample.  The systematics out to $k=0.2$ should therefore be small enough for use in cosmological analyses.  In constrast, the FOG-compressed spectra deviate from a constant bias at the 19\%, 20\% and 30\% levels for the NEAR, MID, and FAR samples; this is a factor of $\sim 5$ larger than the statistical errors on the bandpowers in \citet{tegmark/etal:2006} between $k=0.1$ and $k=0.2$.  The nonlinear correction must therefore be extremely well-calibrated to extract any cosmological information in this regime.\\

We have shown that the nonlinear correction between the reconstructed halo density field and the matter power spectrum is very small.  The small corrections we have found are well fit away from $k=0$ by a polynomial.  We list best fit polynomial parameters in Table~\ref{table:nonlinearfits} in Appendix B.  The amplitude of this correction in our fiducial cosmology should provide a conservative estimate of the uncertainty in the correction as the cosmology is varied within the space allowed by the latest WMAP analysis \citet{komatsu/etal:2008}: $\sim 1\%$ for $k \leq 0.1$ and $\sim 4\%$ between $k=0.1$ and $k=0.2$.  Finally, to recover the linear power spectrum, one must know the amplitude of the BAO suppression and the degree of nonlinearity in the matter power spectrum as a function of the cosmological parameters.  Several groups are addressing this issue \citep[e.g., ][]{eisenstein/seo/white:2007, habib/etal:2007}.\\
\subsubsection{Covariance Matrices}
\begin{figure}
\includegraphics*[scale=0.75,angle=0]{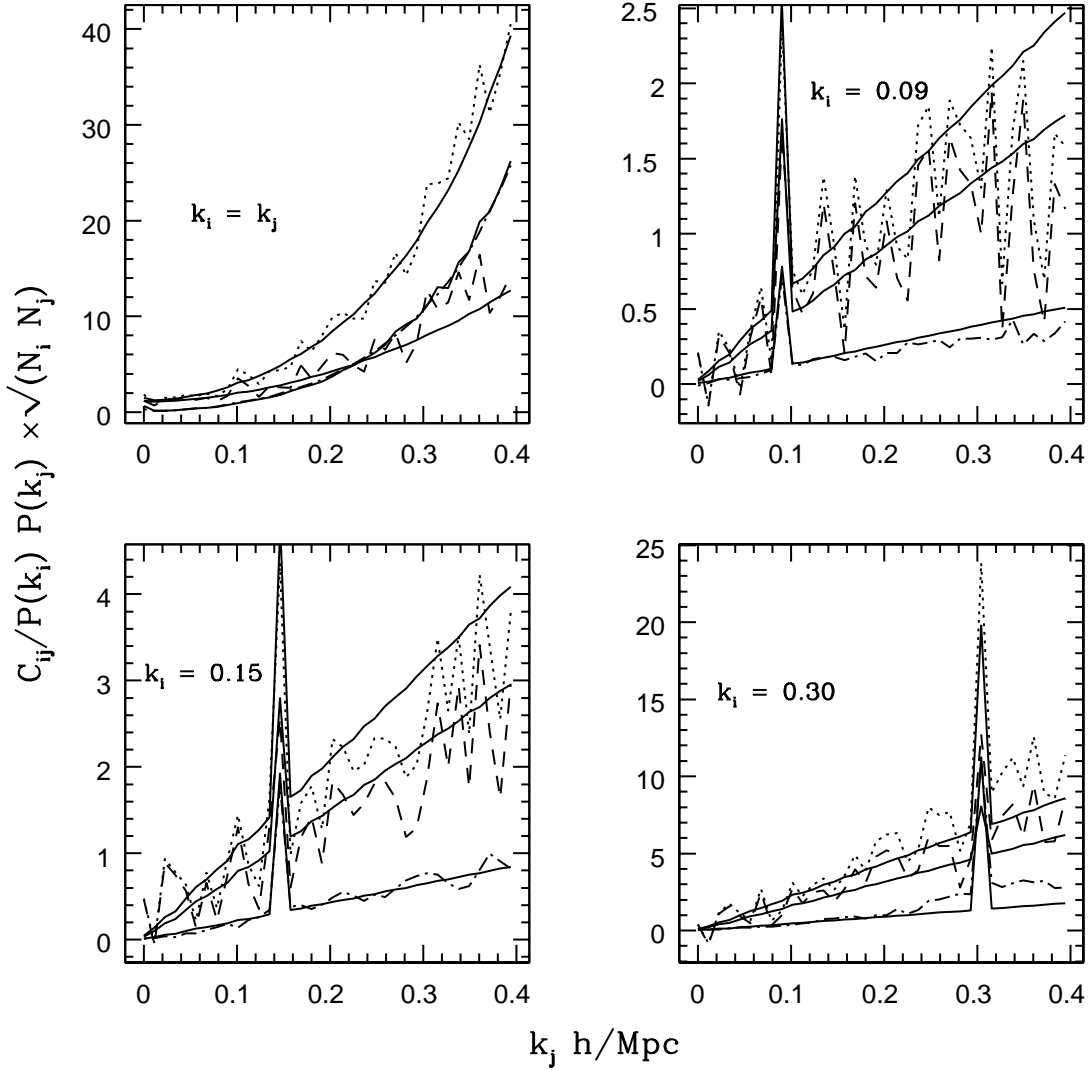}
\caption{\label{fig:covplots} We plot the normalized covariance matrix elements $C_{ij}/P(k_i) P(k_j)$ scaled by $\sqrt{N_i N_j}$, the number modes in bins $i$ and $j$.  We plot our measurements of $C_{ij,tot}$ (dotted curve), $C_{ij,red}$ (dashed curve), and $C_{ij,HOD}$ (dot-dashed curve).  The solid lines show the model fits from Table~\ref{table:covfit1}.  The top left panel shows the diagonal elements, and the other panels show cross sections of the covariance matrix with $k_i =0.09$, 0.15, and 0.3 $h \; {\rm Mpc}^{-1}$.}
\end{figure}

We use the scatter in the power spectrum measurements in the simulations to estimate the covariance matrix for our mock catalogs.  We fit the measurements of the four-point function to a physically motivated model:
\begin{eqnarray}
B_{ij} & = & 4 R_{\alpha} P(k_i)P(k_j)\delta_{DC}^2(z) \label{Bij}\\
C_{ij,tot} & = &\frac{1}{N_{i}} \left(P(k_i)P(k_j) + \frac{\alpha_{tot}}{\bar{n}}\right)^2 \delta^{K}_{ij} + \beta_{tot}B_{ij} \label{cijtotmock}\\
C_{ij,red} & = & \frac{1}{N_{i}} \left(P^2(k_i) + \frac{\alpha_{red}P(k_i)}{\bar{n}} + \frac{1}{\bar{n}^2 N_{mocks}}\right) \delta^{K}_{ij} + \beta_{red}B_{ij} \label{cijredmock} \\
C_{ij,HOD} & = & \frac{1}{N_{i}} \left(\frac{\alpha_{HOD}P(k_i)}{\bar{n}} + \frac{N_{mocks}}{\bar{n}^2 (N_{mocks} - 1)}\right) \delta^{K}_{ij} + \beta_{HOD} B_{ij} . \label{cijhodmock}
\end{eqnarray}
Note we use the measured $P(k_i)$ rather than the linear or model $P(k_i)$ in Eqns.~\ref{Bij}-\ref{cijhodmock}.  In the reduced covariance we expect the Poisson noise to be suppressed by $1/N_{mocks}$, the number of mock catalogs we produce for each TPM simulation.  Eqn.~\ref{cijsum} implies that $\alpha_{HOD} + \alpha_{red} = 2\alpha_{tot}$ for $\alpha_{tot} \approx 1$ and $\beta_{HOD} + \beta_{red} = \beta_{tot}$.  Best fit values for each mock catalog type and each redshift subsample are listed in Table~\ref{table:covfit1}, where the fit included all modes with $k \leq 0.4$, and in Table~\ref{table:covfit2}, where the fit included all modes between $k=0.05$ and $k=0.2$.  We present both in order to assess whether the non-Gaussian terms are growing with $k$. In Fig.~\ref{fig:covplots} we show the measured and model $C_{ij}$ values along the diagonal as well as three rows in the matrix for the MID reconstructed halo density field covariance matrix; the agreement with the model for other rows and redshift subsamples are similar.\\

We find $\alpha_{tot} \approx 1$ for all mocks and redshift subsamples, indicating that the standard shot noise contribution to the covariance is approximately correct for our mock catalogs.  However, the relative distribution of this cross term between $C_{ij,HOD}$ and $C_{ij,red}$ varies; $\alpha_{HOD}$ is large for the FAR sample, where the number density is $\sim$ half that in the MID and NEAR sample, and is also larger in the samples including satellites relative to the samples with only central objects.  All three matrices have significant off-diagonal terms.  $\beta_{tot}$ is larger than unity for all samples, which may be expected since the LRGs are biased tracers, so the DC mode variance will increase by the factor $b^2_{LRG}$.  However, the amplitude does not scale with $b^2$, but is much larger for the FAR sample.  The best fit parameters do not change between real and redshift space for the same sample, but $\beta_{tot}$ is larger for the samples including satellites.  The reconstructed halo density field best fit parameters are consistent with the parameters of the sample of central galaxies, and the FOG-compressed sample is consistent with the sample of central and satellite galaxies.  $\beta_{tot}$ decreases slightly when fitting only to the $k=0.05$ to $k=0.2$ results, but the effect is small and so we adopt the parameters reported in Table~\ref{table:covfit2} to carry out our covariance matrix calculations for the error bars on $P_{LRG}(k)$.\\

There is a modest decrease in the normalized diagonal covariance $C_{ii}/\bar{P}(k_i)^2$ ($\sim 5\%$) when satellite galaxies are included.  However, this decrease is offset by the fact that the $P^2(k_i)$ has a larger nonlinear component, so not all of the available information will be about the linear component; furthermore, the large nonlinear correction introduces uncertainty in extracting the linear component.  Moreover, $\beta_{tot}$ is larger for samples which include satellite galaxies, so the bandpowers will be more highly correlated for those mocks.  We therefore conclude that the reduction in large scale bias by eliminating the satellite galaxies and reconstructing the halo density field is offset in the error budget by smaller off-diagonal covariance and smaller uncertainties in the nonlinear correction.\\ 

\subsubsection{The Redshift Space Quadrupole}
\begin{figure}
\includegraphics*[scale=0.75,angle=0]{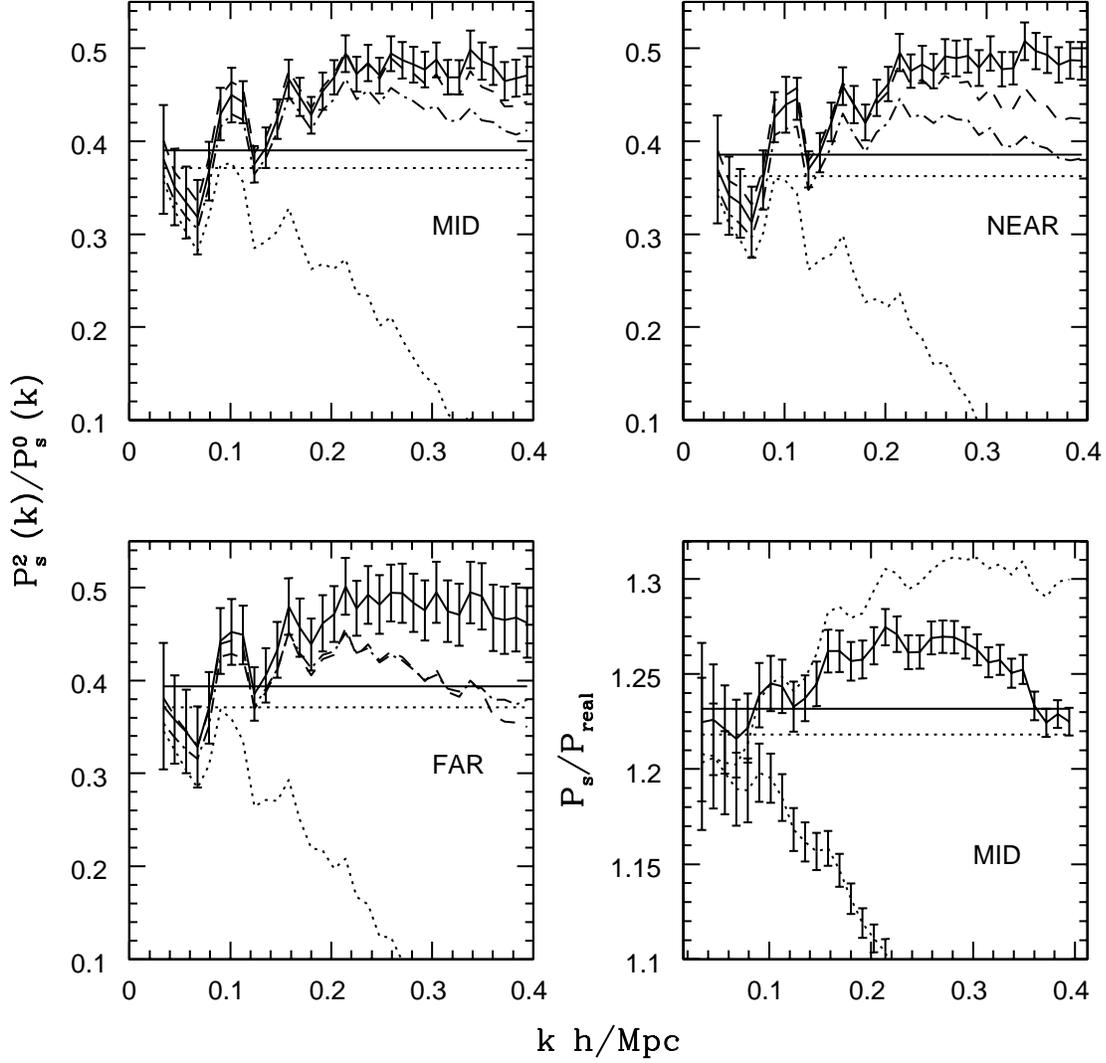}
\caption{\label{fig:quadall}  The bottom right panel shows the ratio of the redshift space monopole to the real space power spectrum for the MID sample.  The straight lines show the prediction of Eqn.~\ref{pseqn} for both samples.  We fit the bias in Eqn.~\ref{betaeqnch4} to the large scale ratio $P_{LRG}/P_{DM}$.  The solid curve is for the sample of central galaxies only ($b_{cen} = 2.11$) and the dotted curve with error bars includes both central and satellite galaxies ($b_{sat} = 2.23$).  The upper dotted curve shows the ratio of the redshift space monopole of the central and satellite galaxies to the real space power spectrum of the central objects (scaled by $b^2_{cen}/b^2_{sat}$).  The other three panels show the quadrupole to monopole ratio in redshift space.  Error bars are only shown for the central galaxy sample (solid curve) for clarity.  The central and satellite sample is shown by the dotted curve, the reconstructed halo density field as the dashed curve, and the FOG-compressed sample as the dash-dot curve.  The expected values from Eqn.~\ref{pseqn} are shown as the straight lines for the central (solid) and central and satellite (dotted) samples.}
\end{figure}

In linear theory the redshift space power spectrum can be decomposed into a monopole, quadrupole, and hexadecapole:
\begin{eqnarray}
\beta & = &\frac{1}{b_{gal}} \frac{d \ln D}{d \ln a}\label{betaeqnch4} \\
P_{s}({\bf k}) & = & P(k) \left[ (1+\frac{2}{3} \beta + \frac{1}{5} \beta^2) L_{0}(\mu_{{\bf k}}) + (\frac{4}{3} \beta + \frac{4}{7} \beta^2) L_{2}(\mu_{{\bf k}}) + (\frac{8}{35} \beta^2) L_{4}(\mu_{{\bf k}})\right]\label{pseqn}
\end{eqnarray}
where $L_i$ is the Legendre polynomial of order $i$, and $\mu_{\bf k} = \hat{{\bf k}} \cdot \hat{{\bf s}}$, with $\hat{{\bf s}}$ the direction along the LOS.  The method of \citet{tegmark/etal:2006} assumes Eqn.~\ref{pseqn} with $\beta$ independent of $k$ to obtain the best estimate of the real space power spectrum from $P_{gg}$, $P_{gv}$, and $P_{vv}$.  In Fig.~\ref{fig:quadall} we examine the structure of the redshift space distortions in our mock catalogs.  The first three panels show the quadrupole to monopole ratio for central LRGs (solid), the reconstructed halo density field (dashed), and the FOG-compressed density field (dash-dot).  These are all similar and show a modest increase in the quadrupole to monopole ratio with $k$.  This demonstrates that the FOG compression technique of \citet{tegmark/etal:2006} successfully removes the effects of the FOGs induced by satellite galaxies.  The oscillations probably result from the extra suppression of BAO features in redshift space \citep{eisenstein/seo/white:2007}.  The central and satellite LRG sample (dotted) quadrupole to monopole ratio lies below the expected value at all $k$ values, indicating that the FOG suppression of power is evident even in the linear regime.  Therefore a sample where the effects of satellite FOGs have not been removed by either halo density field reconstruction or FOG compression will provide a biased estimate of $\beta$.\\

The bottom right panel shows the ratio of the redshift space monopole to the real space power spectrum for the sample of central galaxies only (solid) and central and satellite galaxies (dotted).  The suppression of power by the satellite FOGs is also evident here.  For central LRGs only, the ratio is consistent with the expected value from Eqn.~\ref{pseqn}, and between $k=0$ and $k=0.2$ this ratio varies by only $\sim 3\%$.  While there is a clear scale dependence in the quadrupole to monopole ratio for the central LRG sample, the large bias of LRGs means that the monopole redshift space power spectrum is nearly insensitive to the redshift space nonlinearities in the halo density field.  The upper dotted curve shows $P_{s,sat}/P_{real,cen}\times b^2_{cen}/b^2_{sat}$.  The increase in power in real space when the satellites are included is larger than the suppression of power in redshift space by their FOGs, so that the satellite monopole spectrum has more power at high $k$ than the real space central LRGs.\\ 

\section{Conclusions}
This paper introduced and tested our algorithm for using the halo density field to estimate the underlying matter power spectrum.  We found the nonlinear correction between this field and the underlying matter density field to be both smaller and more robust to variations in the effects of satellite galaxies than both the FOG-compressed density field used in the analysis of \citet{tegmark/etal:2006} and the redshift space monopole power spectrum used in the analysis of \citet{percival/etal:2007}.\\

The parameters of our simulation set were selected to provide accurate two point halo density and velocity statistics to assure accurate representations of FOG features in our mock catalogs.  To our knowledge, this is the first attempt to study the detailed effects of FOG treatment on the LRG power spectrum.  We find that the FOG treatment can affect bandpowers even in the linear regime, $k \leq 0.1$.
We first examined the nonlinear matter power spectrum of our 42 simulations.  The ratio of the nonlinear to input matter power spectra is well described by a smearing of the BAOs as modeled by \citet{eisenstein/seo/white:2007} and a smooth increase in power with $k$ that can be fit by a third order polynomial out to $k =0.4 \; h \; {\rm Mpc}^{-1}$ or second order polynomial out to $k=0.2 \; h \; {\rm Mpc}^{-1}$.  We detect a substantial deviation from the predictions of halofit \citep{smith/etal:2003}.\\

Fig.~\ref{fig:LRGratMIDall} demonstrates the main point of this work: satellite galaxies systematically alter the shape of the power spectrum {\em even} at $k < 0.1 \; h \; {\rm Mpc}^{-1}$.  Extraction of cosmological information from the broadband shape of the power spectrum is already limited by systematics \citep{sanchez/cole:2007} which we (and others) suggest can be attributed primarily to differences in the satellite contribution to the power spectrum.  In this paper we demonstrate that while the FOG compression scheme in \citet{tegmark/etal:2006} exacerbates these issues and requires a large nonlinear correction, the FOG features in the density field can be used to reconstruct the halo density field with high fidelity.  The power spectrum of this field deviates from the dark matter power spectrum at the $\leq 4\%$ level for $k \leq 0.2 \; h \; {\rm Mpc}^{-1}$ and $\leq 1\%$ level for $k \leq 0.1 \; h \; {\rm Mpc}^{-1}$, where cosmological analyses usually restrict themselves.  Moreover, we have shown that this correction changes only slightly between the NEAR, MID, and FAR LRG reconstructed halo density fields, while the FOG compressed mocks have a much larger variation between samples.  Therefore, we can hope to push cosmological analyses to larger $k$ using the reconstructed halo density field as a tracer of the underlying matter density field fluctuations--- particularly for galaxy samples like the LRGs which are spread over a large redshift range and are not volume-limited, and thus have substantial variation in the satellite contribution to the power spectrum with redshift.\\

While we have not addressed the variation of the nonlinear correction to the reconstructed halo density field as a function of cosmological parameters, we have designed the form of our correction to minimize the variation with cosmology.  Other researchers \citep[e.g., ][]{habib/etal:2007} are studying the dark matter power spectrum as a function of cosmology.  We expect that our nonlinear correction $P_{LRG}/P_{DM}$ will remain small (of order 4\% below $k = 0.2 \; h \; {\rm Mpc}^{-1}$) as the cosmology is varied, so the variation of this small correction should be even smaller.  Therefore, instead of introducing a nuisance parameter for the nonlinear correction, we propose that the amplitude of the correction should be taken as the error on its value in cosmological parameter analyses, or relatively strong priors on the amplitude of the correction to the reconstructed halo density field be introduced.\\

In this work we have also investigated the properties of the power spectrum covariance matrix for the dark matter, as well as the mock galaxy catalogs divided into 6 different samples: central galaxies in real and redshift space, central and satellite galaxies in real and redshift space, our reconstructed halo density field in redshift space, and the \citet{tegmark/etal:2006} FOG compressed galaxy density field in redshift space.  All of these samples were well-modeled by a diagonal matrix with the usual Gaussian and Poisson shot noise terms plus the beat-coupling term presented in \citet{hamilton/rimes/scoccimarro:2006}.  We expect that this will be a useful model for fitting the survey covariance matrix, where $P(k_b)$ is replaced by $P(k_{survey})$ with $k_{survey}$ determined by some effective survey size.\\

Finally we examined the structure of the redshift space distortions as a function of $k$ using the quadrupole.  Both the reconstructed halo density field and the FOG-compressed mock LRG samples reproduce the modest $k$ dependence of the halo density field quadrupole to monopole ratio.  Since the LRGs are so highly biased, this scale dependence causes a $\lesssim 3\%$ deviation in the redshift space monopole to real space power spectrum ratio out to $k=0.2$.  When satellites are included in the sample without FOG compression, the quadrupole to monopole ratio is lower than the linear value for the entire $k$ range accessible in our simulations and will significantly bias the estimate of $\beta$.\\

\section{Acknowledgments}
We thank Daniel Eisenstein for providing our SDSS LRG sample and Michael Blanton for providing the inverse random catalogs and Tycho2 catalog.  We thank Jeremy Tinker for providing the SO halo-finding code used to produce our halo catalogs.  We thank Will Percival and Licia Verde for excellent discussions.\\

Funding for the Sloan Digital Sky Survey (SDSS) has been provided by the Alfred P. Sloan Foundation, the Participating Institutions, the National Aeronautics and Space Administration, the National Science Foundation, the U.S. Department of Energy, the Japanese Monbukagakusho, and the Max Planck Society. The SDSS Web site is http://www.sdss.org/.\\

The SDSS is managed by the Astrophysical Research Consortium (ARC) for the Participating Institutions. The Participating Institutions are The University of Chicago, Fermilab, the Institute for Advanced Study, the Japan Participation Group, The Johns Hopkins University, Los Alamos National Laboratory, the Max-Planck-Institute for Astronomy (MPIA), the Max-Planck-Institute for Astrophysics (MPA), New Mexico State University, University of Pittsburgh, Princeton University, the United States Naval Observatory, and the University of Washington.\\

Computer simulations were supported by the
National Science Foundation through TeraGrid resources
provided by Pittsburgh Supercomputing Center
and the National Center for Supercomputing Applications
under grant AST070021N; simulations were also
performed at the TIGRESS high performance
computer center at Princeton University, which is jointly supported by
the Princeton Institute for Computational Science and Engineering and
the Princeton University Office of Information Technology.\\

This work draws on Chapters 3 and 4 of BAR's Ph.D. thesis in Princeton University Department of Physics.  BAR gratefully acknowledges support from both the National Science Foundation Graduate Research Fellowship and PIRE program.  This project was supported by NSF Grant 0707731.  DNS thanks the APC (Universite de Paris VII) for its hospitality while this work was completed.\\
\bibliographystyle{apj}
\bibliography{apj-jour,/Users/breid/bibfile/bethbibfile}
\section{Appendix A: Resolution Study Results}
The goal of the resolution study detailed in Chapter 3 of \citet{reid:phd} was to determine the minimum $N$-body simulation mass resolution satisfying two criteria.
\begin{itemize}
\item The minimum halo mass is small enough to accommodate broad $N_{cen}(M)$ functions (i.e., large values of $\sigma_{log M}$).  The  width of this function is dictated by the observed large-scale clustering amplitude, either in $P_{LRG}(k)$ or $w_p(r_p)$, which determines $\sigma_8 b_{LRG}$.
\item The finite mass resolution does not systematically bias the small scale $n$-point halo statistics that must be accurate in order to test our method of halo density field reconstruction.  We argue in \citet{reid:phd} that agreement of several two-point clustering and velocity statistics and the CiC group multiplicity functions is sufficient for accurate mock FOG structures.  Furthermore, the small observed deviations in the halo mass function at low $M$ can be absorbed by the parameters of $N_{cen}(M)$.
\end{itemize}
For the cosmological parameters of our mock catalogs and the observed LRG clustering strength, the first condition places the more stringent requirement on the simulation mass resolution.  To establish the level of systematics relevant to the second condition, we compare the results of one $N$-body simulation with our adopted parameters $L_{box} = 558$ $h^{-1}$ Mpc and $N_{p,med} = 512^3$ ($M_{p,med} = 1.43 \times 10^{11} M_{\sun}$) with a higher resolution simulation ($N_{p,high} = 640^3$, $M_{p,high} = 7.33 \times 10^{10} M_{\sun}$) and a lower resolution simulation ($N_{p,low} = 384^3$, $M_{p,low} = 3.39 \times 10^{11} M_{\sun}$) of the same initial conditions.  We highlight the findings of this comparison below for the simulation outputs near our $z_{MID} = 0.342$, and refer the interested reader to \citet{reid:phd} for a more thorough discussion.\\
\subsection{Resolution Effects in the Halo Mass Function and the Halo Occupation Distribution}
The mass functions from the low, medium, and high resolution simulations are in excellent agreement with one another and within a few percent of the \citet{tinker/etal:2008} mass function.  There is a slight increase ($\sim 5\%$) in the halo mass function as the aggressive mass limit $50 M_p$ of our SO halo catalogs is reached; this should be kept in mind in future analyses, but is not inherently problematic since slight changes in the mass function in this regime are degenerate with small variations in the form of $N_{cen}(M)$.\\

We produce LRG mock catalogs at $\sigma_{log M} = 0.6$, the value that matches the observed LRG clustering amplitude, in the low, medium, and high resolution simulations.  For the high resolution simulation, only 2.6\% of the mock LRGs occupy halos below the halo catalog mass limit of the medium resolution simulation, and only 11\% are below $100M_{p,med}$.  Therefore, the abrupt cutoff in $N_{cen}(M)$ induced by our halo catalog mass limit will show minimal differences with a mock catalog derived from a halo catalog extending to a substantially lower mass limit.  In addition, even if there are slight systematics in the pair statistics of a halo catalog in the $\sim 50-100 M_{p,med}$ mass range, they will be suppressed by at least a factor of $\sim 10$.\\

\subsection{Two-point clustering and velocity statistics}
There is a slight but systematic decrease in power with $k$ as the simulation resolution decreases.  This is a $0.2\%$ effect at $k=0.2$ $h \; {\rm Mpc}^{-1}$ between the medium and high resolution simulations, well within the statistical errors and modeling uncertainties between the LRGs and dark matter density fields.\\

\begin{figure}
\includegraphics*[scale=0.75,angle=0]{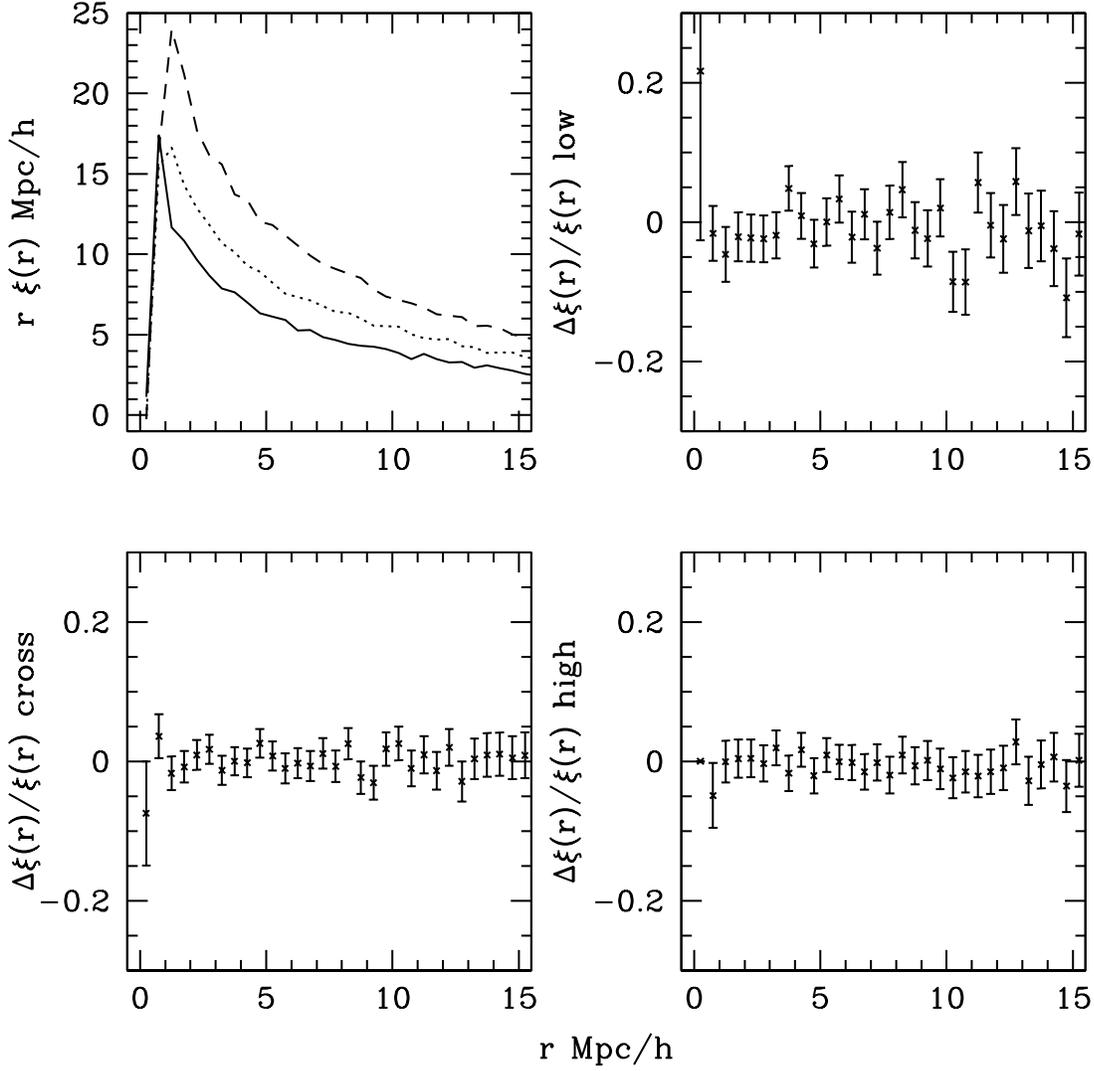}
\caption{\label{fig:dxiplot} {\it Upper left:} $r \xi_{ij}(r)$ on small scales for the low mass halos auto-correlation (solid), high mass halos auto-correlation (dashed), and their cross correlation (dotted).  Also shown are $(\xi_{high}(r) - \xi_{med}(r))/\xi_{high}(r)$ computed in bins of size $\Delta r = 0.5$ $h^{-1}$ Mpc for the low mass halos auto-correlation (upper right), high mass halos auto-correlation (lower right), and their cross correlation (lower left).  The agreement for each correlation extends to $r = 150$ $h^{-1}$ Mpc (not shown).}
\end{figure}

\begin{figure}
\includegraphics*[scale=0.75,angle=0]{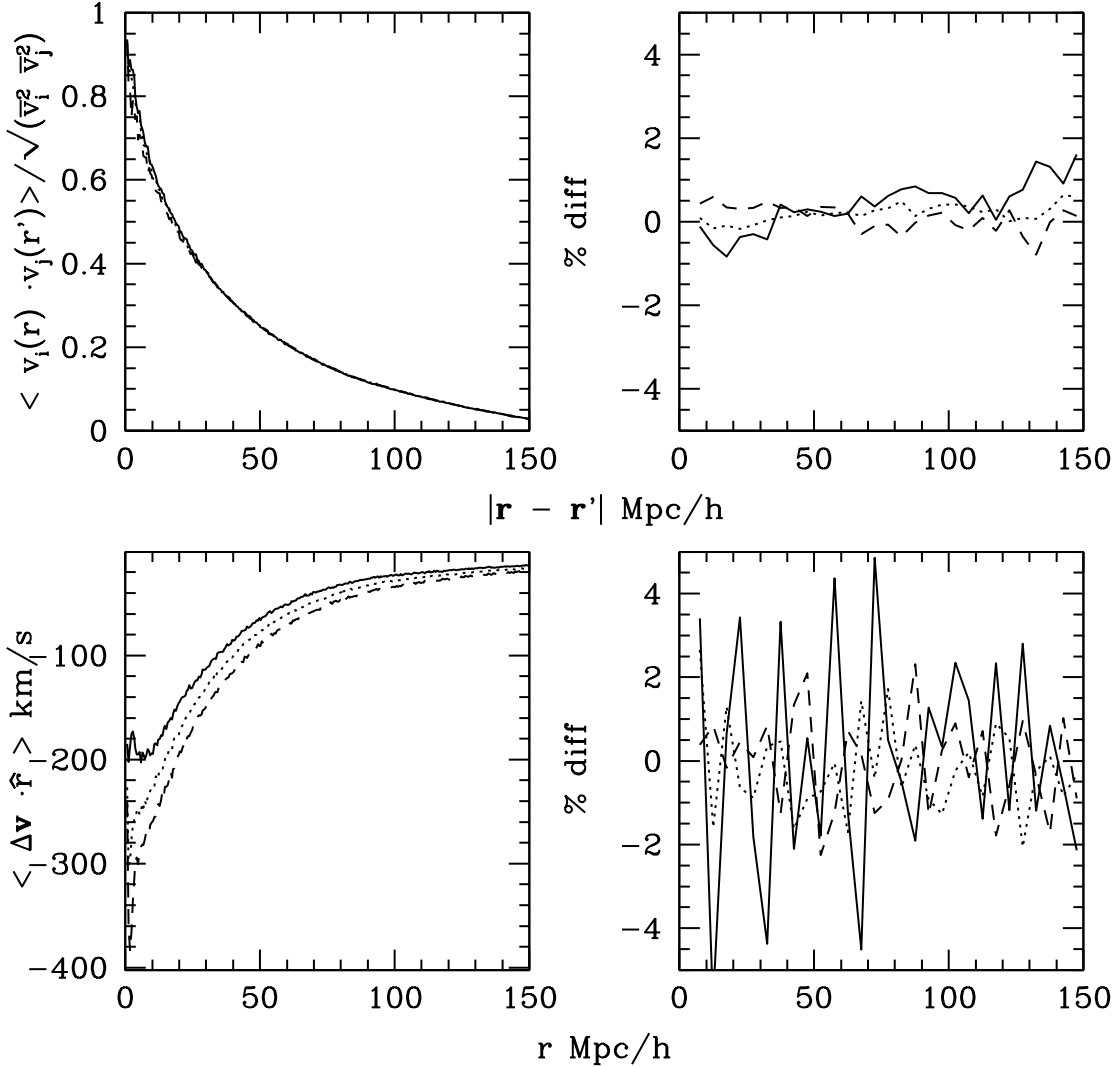}
\caption{\label{fig:dvcorrplot} {\it Upper left:} The normalized halo-halo velocity correlation function, $\left<{\bf v}_i({\bf r}) \cdot {\bf v}_j({\bf r}')\right>/\sqrt{\left<{\bf v}_i^2\right >\left<{\bf v}_j^2\right >}$, vs $|{\bf r} - {\bf r}'|$ for the low mass halos auto-correlation (solid), high mass halos auto-correlation (dashed), and their cross correlation (dotted) binned in intervals of 0.5 $h^{-1}$ Mpc in $|{\bf r} - {\bf r}'|$.  {\it Upper right:} The fractional difference in the normalized halo-halo velocity correlation between the high and medium resolution simulations binned in intervals of 5 $h^{-1}$ Mpc in $|{\bf r} - {\bf r}'|$.  {\it Lower panels:} Same as the upper panels, but for the halo-halo relative velocity correlation function along the halo separation vector, $\left<\left({\bf v}_i({\bf r}) - {\bf v}_j({\bf r}')\right) \cdot ({\bf r} - {\bf r}')/|{\bf r} - {\bf r}'|\right>$.}
\end{figure}

We split the halos into two mass bins: $50-99 M_{p,med}$ and $M \ge 100 M_{p,med}$.  This corresponds to $98-195$ particles per halo for the low mass bin of the high resolution simulation.  We did not detect any significant resolution-dependent differences in $P(k)$ and $\xi(r)$, other than a $\lesssim 1\%$ change in the large scale bias of the halo samples with respect to the matter.  Fig.~\ref{fig:dxiplot} shows that the difference between the medium and high resolution simulation $\xi_{ij}(r)$, for both auto- and cross-correlations of the low and high mass halo bins, is consistent with 0 at small scales ($r \lesssim 15$ $h^{-1}$ Mpc); this is true out to 150 $h^{-1}$ Mpc.  Fig.~\ref{fig:dvcorrplot} shows the velocity correlation statistic $\left<{\bf v}_i({\bf r}) \cdot {\bf v}_j({\bf r}')\right>/\sqrt{\left<{\bf v}_i^2\right >\left<{\bf v}_j^2\right >}$ for halo mass bins $i$ and $j$.  We find $\leq 1\%$ level agreement between the medium and high resolution simulations.  In Fig.~\ref{fig:dvcorrplot}  we also examine the relative halo velocities along the vector separating the halos: $\left<\left({\bf v}_i({\bf r}) - {\bf v}_j({\bf r}')\right) \cdot ({\bf r} - {\bf r}')/|{\bf r} - {\bf r}'|\right>$ and find no evidence for systematic bias between the medium and high resolution simulations.\\

\subsection{Mock catalog statistics: two point clustering statistics and multiplicity bias $b(n_{sat})$}
In \citet{reid:phd} we compare both the power spectrum $P(k)$ and the projected correlation function $w_p(r_p)$ of the medium and high resolution simulations, for which we find systematic deviations smaller than $\sim 1/6$ the statistical error induced by the stochasticity of the halo occupation.
We quantify the difference between the true one-halo group density field and the CiC group density field by the multiplicity bias $b(n_{sat}) = N_{CiC}(n_{sat})/N_{true}(n_{sat})$.  We show that $b(n_{sat})$ systematically varies by less than the amount of variation induced by the halo occupation stochasticity for mock catalogs with $\sigma_{log M} = 0.8$, which pushes to lower mass limits than our better matched $\sigma_{log M} = 0.6$ catalogs.\\

We conclude that the selected mass resolution of our simulations is well-optimized to produce a large simulation volume over several realizations and with well-controlled systematic biases induced by the finite mass resolution.\\

\section{Appendix B: Power Spectra and Covariance Matrix Best Fit Parameters}
\begin{deluxetable}{lllllllll}
\tabletypesize{\scriptsize}
\tablewidth{0pt}
\tablecolumns{9}
\tablecaption{\label{table:nonlinearfits} Fits to nonlinear power spectra.  $k_{max,fit}$ and $k_{BAO}$ are in units of $h \; {\rm Mpc}^{-1}$.  For $P_{DM}(k)/P_{IC}(k)$ we fit to Eqn.~\ref{Pnlmat}.  When $k_{max,fit} = 0.2$ we hold $a_3 = 0$.  Some of the redshift space mock catalog fits are not well-behaved at very small $k$'s, and the corrections to a constant should be suppressed at $k \lesssim 0.05$.  This also makes $a_0$ deviate from the large scale bias, given in the last column and fit using only $k < 0.06$.  $P_{sat,z}(k)$ denotes the power spectrum of a catalog containing central and satellite galaxies in redshift space, $P_{CiC,z}(k)$ denotes the power spectrum of the reconstructed halo density field in redshift space, and $P_{teg,z}(k)$ denotes an FOG compressed following \citet{tegmark/etal:2006}.}
\tablehead{
\colhead{sample} & \colhead{function} & \colhead{$k_{max,fit}$} & \colhead{$k_{BAO}$} & \colhead{$a_0$} & \colhead{$a_1$} & \colhead{$a_2$} & \colhead{$a_3$} & \colhead{$b^2_{LRG}$}}
\startdata
NEAR & $P_{DM}(k)/P_{IC}(k)$ & 0.2 & 0.12 & 1.025 & -1.109 & 9.296 & -- & -- \\
NEAR & $P_{DM}(k)/P_{IC}(k)$ & 0.4 & 0.13 & 1.027 & -1.305 & 12.05 & -9.136 & -- \\
NEAR & $P_{sat,z}(k)/P_{DM}(k)$ & 0.2 & -- & 5.14 & 5.38 & -15.96 & -- & 5.38 \\
NEAR & $P_{sat,z}(k)/P_{DM}(k)$ & 0.4 & -- & 5.01 & 7.91 & -28.72 & 15.08 & 5.38 \\
NEAR & $P_{CiC,z}(k)/P_{DM}(k)$ & 0.2 & -- & 4.74 & 2.11 & -12.80 & -- & 4.83 \\
NEAR & $P_{CiC,z}(k)/P_{DM}(k)$ & 0.4 & -- & 4.66 & 3.68 & -20.45 & 8.40 & 4.83 \\
NEAR & $P_{teg,z}(k)/P_{DM}(k)$ & 0.2 & -- & 5.15 & 4.85 & 7.78 & -- & 5.42 \\
NEAR & $P_{teg,z}(k)/P_{DM}(k)$ & 0.4 & -- & 5.03 & 6.44 & 6.41 & -20.33 & 5.42 \\
MID & $P_{DM}(k)/P_{IC}(k)$ & 0.2 & 0.13 & 1.025 & -1.042 & 8.614 & -- & -- \\
MID & $P_{DM}(k)/P_{IC}(k)$ & 0.4 & 0.14 & 1.026 & -1.199 & 11.06 & -8.426 & -- \\
MID & $P_{sat,z}(k)/P_{DM}(k)$ & 0.2 & -- & 5.67 & 6.89 & -17.9 & -- & 5.99 \\
MID & $P_{sat,z}(k)/P_{DM}(k)$ & 0.4 & -- & 5.54 & 9.21 & -28.3 & 8.54 & 5.99 \\
MID & $P_{CiC}(k)/P_{DM}(k)$ & 0.2 & -- & 5.30 & 3.43 & -14.7 & -- & 5.46 \\
MID & $P_{CiC}(k)/P_{DM}(k)$ & 0.4 & -- & 5.21 & 5.13 & -21.8 & 4.35 & 5.46 \\
MID & $P_{teg}(k)/P_{DM}(k)$ & 0.2 & -- & 5.68 & 6.11 & 8.23  & -- & 6.03 \\
MID & $P_{teg}(k)/P_{DM}(k)$ & 0.4 & -- & 5.58 & 7.27 & 10.4 & -28.4 & 6.03 \\
FAR & $P_{DM}(k)/P_{IC}(k)$ & 0.2 & 0.13 & 1.025 & -0.994 & 8.144 & -- & -- \\
FAR & $P_{DM}(k)/P_{IC}(k)$ & 0.4 & 0.15 & 1.025 & -1.126 & 10.39 & -7.945 & -- \\
FAR & $P_{sat,z}(k)/P_{DM}(k)$ & 0.2 & -- & 6.12 & 9.72 & -19.26 & -- & 6.58 \\
FAR & $P_{sat,z}(k)/P_{DM}(k)$ & 0.4 & -- & 5.89 & 13.96 & -37.46 & 11.39 & 6.58 \\
FAR & $P_{CiC}(k)/P_{DM}(k)$ & 0.2 & -- & 5.79 & 4.36 & -13.97 & -- & 6.00 \\
FAR & $P_{CiC}(k)/P_{DM}(k)$ & 0.4 & -- & 5.64 & 6.94 & -23.48 & -- & 6.00 \\
FAR & $P_{teg}(k)/P_{DM}(k)$ & 0.2 & -- & 6.16 & 7.90 & 26.43 & -- & 6.63 \\
FAR & $P_{teg}(k)/P_{DM}(k)$ & 0.4 & -- & 5.96 & 10.49 & 26.82 & -45.03 & 6.63 \\ 
\enddata
\end{deluxetable}
\begin{deluxetable}{lllllll}
\tabletypesize{\scriptsize}
\tablewidth{0pt}
\tablecolumns{7}
\tablecaption{\label{table:covfit1} Fits for the mock catalog covariance matrices using Eqns.~\ref{Bij} - ~\ref{cijhodmock}; all bandpowers between $k=0.01$ and $k=0.4$ are included in the fits.  `cen, real/redshift' denotes a sample of only central LRGs in real/redshift space; `sat, real/redshift' denotes a sample with both central and satellite galaxies in real/redshift space; `CiC, redshift' denotes the reconstructed halo density field using the CiC method, and `Teg, redshift' denotes the FOG-compressed sample according to \citet{tegmark/etal:2006}.}
\tablehead{
\colhead{Catalog} & \colhead{$\alpha_{HOD}$} & \colhead{$\alpha_{red}$} & \colhead{$2\alpha_{tot}$} & \colhead{$\beta_{HOD}$} &\colhead{$\beta_{red}$} & \colhead{$\beta_{tot}$}}
\startdata 
NEAR cen, real & 1.07 & 1.10 & 2.00 & 0.26 & 0.94 & 1.26 \\
NEAR sat, real & 1.30 & 0.89 & 1.98 & 0.36 & 1.58 & 1.97 \\
NEAR cen, redshift & 1.11 & 0.99 & 2.02 & 0.26 & 0.95 & 1.29  \\
NEAR sat, redshift & 1.27 & 0.63 & 2.01 & 0.29 & 1.28 & 1.64 \\
NEAR CiC, redshift & 1.11 & 0.92 & 2.02 & 0.26 & 0.94 & 1.30 \\
NEAR Teg, redshift & 1.39 & 0.74 & 1.98 & 0.35 & 1.60 & 1.99 \\
MID cen, real & 1.14 & 1.0 & 2.00 & 0.34 & 1.21 & 1.62 \\
MID sat, real & 1.32 & 0.77 & 1.99 & 0.42 & 1.71 & 2.20 \\
MID cen, redshift & 1.17 & 0.93 & 2.01 & 0.34 & 1.19 & 1.63 \\
MID sat, redshift & 1.30 & 0.61 & 2.00 & 0.36 & 1.47 & 1.92 \\
MID CiC, redshift & 1.16 & 0.86 & 2.01 & 0.34 & 1.18 & 1.63 \\
MID Teg, redshift & 1.41 & 0.58 & 1.97 & 0.41 & 1.73 & 2.20 \\
FAR cen, real & 1.65 & 0.50 & 2.05 & 1.32 & 1.62 & 3.13 \\
FAR sat, real & 1.85 & 0.38 & 2.07 & 1.71 & 2.67 & 4.50 \\
FAR cen, redshift & 1.66 & 0.42 & 2.05 & 1.25 & 1.64 & 3.13 \\
FAR sat, redshift & 1.79 & 0.16 & 2.07 & 1.37 & 2.08 & 3.70 \\
FAR CiC, redshift & 1.66 & 0.37 & 2.06 & 1.25 & 1.69 & 3.21 \\
FAR Teg, redshift & 1.90 & 0.13 & 2.06 & 1.54 & 2.53 & 4.19 \\
\enddata
\end{deluxetable}
\begin{deluxetable}{lllllll}
\tabletypesize{\scriptsize}
\tablewidth{0pt}
\tablecolumns{7}
\tablecaption{\label{table:covfit2}  Fits for the mock catalog covariance matrices using Eqns.~\ref{Bij} - ~\ref{cijhodmock}; all bandpowers between $k=0.05$ and $k=0.2$ are included in the fits.  Samples are the same as in Table~\ref{table:covfit1}.}
\tablehead{
\colhead{Catalog} & \colhead{$\alpha_{HOD}$} & \colhead{$\alpha_{red}$} & \colhead{$2\alpha_{tot}$} & \colhead{$\beta_{HOD}$} &\colhead{$\beta_{red}$} & \colhead{$\beta_{tot}$}}
\startdata 
NEAR cen, real & 1.05 & 0.85 & 1.96 & 0.23 & 0.86 & 1.14 \\
NEAR sat, real &  1.22 & 0.62 & 1.94 & 0.29 & 1.43 & 1.76 \\
NEAR cen, redshift & 1.07 & 0.71 & 1.90 & 0.23 & 0.94 & 1.23 \\
NEAR sat, redshift & 1.22 & 0.51 & 1.86 & 0.26 & 1.33 & 1.66 \\
NEAR CiC, redshift & 1.07 & 0.68 & 1.90 & 0.24 & 0.97 & 1.27 \\
NEAR Teg, redshift & 1.28 & 0.45 & 1.86 & 0.29 & 1.47 & 1.82 \\
MID cen, real & 1.08 & 0.84 & 1.96 & 0.28 & 1.14 & 1.49 \\
MID sat, real & 1.21 & 0.62 & 1.93 & 0.35 & 1.60 & 2.02 \\
MID cen, redshift & 1.11 & 0.74 & 1.93 & 0.30 & 1.20 & 1.58 \\
MID sat, redshift & 1.23 & 0.54 & 1.88 & 0.33 & 1.55 & 1.96 \\
MID CiC, redshift & 1.10 & 0.72 & 1.91 & 0.29 & 1.24 & 1.61 \\
MID Teg, redshift & 1.29 & 0.46 & 1.88 & 0.36 & 1.66 & 2.10 \\
FAR cen, real & 1.57 & 0.36 & 2.00 & 1.08 & 1.66 & 2.87 \\
FAR sat, real & 1.68 & 0.21 & 2.00 & 1.35 & 2.51 & 3.98 \\
FAR cen, redshift & 1.57 & 0.26 & 1.99 & 1.06 & 1.70 & 2.94 \\
FAR sat, redshift & 1.67 & 0.13 & 1.98 & 1.20 & 2.28 & 3.65 \\
FAR CiC, redshift & 1.58 & 0.26 & 2.00 & 1.06 & 1.84 & 3.09 \\
FAR Teg, redshift & 1.73 & 0.06 & 1.99 & 1.31 & 2.47 & 3.93 \\
\enddata
\end{deluxetable}
\end{document}